\documentclass[11pt,a4paper,preprintnumbers,nofootinbib]{revtex4}
\pdfoutput=1

\usepackage[T1]{fontenc}
\usepackage[utf8]{inputenc}

\usepackage[english]{babel}

\usepackage{graphicx}
\usepackage[dvipsnames]{xcolor}

\usepackage{amsmath}
\usepackage{amsfonts}
\usepackage{amssymb}
\usepackage{amstext}
\usepackage{slashed}

\usepackage{multirow}
\usepackage{booktabs}

\usepackage{hyperref}
\usepackage{placeins}
\usepackage{verbatim}

\def \re{\textrm{Re}}
\def \im{\textrm{Im}}

\def \azeL{{H_0^L}}

\def \apaL{{H_\parallel^L}}

\def \apeL{{H_\perp^L}}

\begin{document}

\title{Null tests from angular distributions in $D  \to P_1 P_2  l^+l^-$, $l=e,\mu$ decays  on and off peak}
\author{Stefan de Boer$^{\,a}$}
\email{stefan.boer@kit.edu}
\author{Gudrun Hiller$^{\,b}$}
\email{ghiller@physik.uni-dortmund.de}
\affiliation{$^{\,a}$ Institut f\"ur Theoretische Teilchenphysik, Karlsruher Institut f\"ur  Technologie, D-76128 Karlsruhe, Germany\\
$^{\,b}$ Fakultät  Physik, TU Dortmund, Otto-Hahn-Str.4, D-44221 Dortmund, Germany}

\preprint{DO-TH 18/11, QFET-2018-09, TTP18-018}

\begin{abstract}
We systematically analyze the full angular distribution in $D \to P_1 P_2 l^+ l^-$ decays, where $P_{1,2}=\pi,K$, $l=e,\mu$.
We identify  several null tests of the standard model (SM). Notably, the angular coefficients $I_{5,6,7}$,  driven by  the leptons' axial-vector coupling $C_{10}^{(\prime)}$,
vanish by means of a superior GIM-cancellation and are protected by parity invariance below the weak scale.
CP-odd observables related to the angular coefficients $I_{5,6,8,9}$ allow to measure CP-asymmetries without $D$-tagging.
The corresponding observables $A_{5,6,8,9}$ constitute null tests of the SM. 
Lepton universality in $|\Delta c| =|\Delta u|=1$ transitions can be tested by comparing  $D  \to P_1 P_2  \mu^+ \mu^-$ to $D  \to P_1 P_2  e^+ e^-$ decays. Data for  $P_1 P_2=\pi^+ \pi^-$ and $K^+ K^-$ on muon modes are available from LHCb and on  electron modes from BESIII. Corresponding ratios  of dimuon to dielectron branching fractions are at least about an order of magnitude away from probing  the SM. In the future
 electron and muon measurements should be made available for the same cuts as corresponding ratios $R_{P_1 P_2}^D$ provide null tests of $e$-$\mu$-universality.
We work out  beyond-SM signals  model-independently  and  in SM extensions with  leptoquarks.
\end{abstract}

\maketitle

\section{Introduction}

Rare  charm decays are notoriously challenging theoretically, yet offer singular insights into flavor  in the up-quark sector \cite{Burdman:2001tf}. 
With standard model (SM)  branching ratios of $|\Delta c| =|\Delta u|=1$ modes in the $10^{-7}-10^{-6}$  (semileptonic) and $10^{-6}-10^{-4}$ (radiative) range,  precision studies are feasible at  the experiments LHCb \cite{Bediaga:2012py}, Belle II \cite{Aushev:2010bq} and BESIII \cite{Asner:2008nq}.
In view of the substantial hadronic uncertainties  there are three  main avenues  to probe for beyond the standard model (BSM) physics in charm: 
 {\it i)} a measurement 
in an obvious excess of the SM such as the $D \to \pi \mu^+ \mu^-$ branching ratio at  high dilepton mass  \cite{Aaij:2013sua} -- a window that can be  closing soon \cite{deBoer:2015boa}, 
{\it ii)} extract the  SM contribution from a SM-dominated mode and use $SU(3)_F$, {\it e.g.,} recently  demonstrated for $D \to V \gamma$, $V=\rho, \bar K^{*}, \phi$ and $D_{(s)} \to K \pi \pi \gamma$ decays in \cite{deBoer:2018zhz} 
or {\it iii)}  perform  null tests of (approximate) symmetries of the SM.
The latter includes searches for lepton flavor violation (LFV), CP-violation, or lepton non-universality (LNU).

In this work we consider  angular observables, and LNU tests in semileptonic rare charm decays into electrons and muons.
Exclusive semileptonic 3-body charm decays have been studied in some detail in the decays
$D \to \pi l^+ l^-$ \cite{deBoer:2015boa,Fajfer:2005ke} and $D \to \rho l^+ l^- $ \cite{Fajfer:1998rz,Burdman:2001tf,Fajfer:2005ke} within  QCD factorization (QCDF)  \cite{Feldmann:2017izn}.
Previous theory works on the four-body decays $D \to P_1P_2 l^+ l^-$, $P_{1,2}=\pi,K$ decays  highlight T-odd asymmetries \cite{Bigi:2011em,Cappiello:2012vg}  or  the  leptonic  forward-backward asymmetry  \cite{Cappiello:2012vg}, however, a
systematic analysis of the virtues of the full angular distribution at par with the corresponding one in $B$-decays \cite{Das:2014sra} is  missing.
Modes sensitive to BSM physics in semileptonic transitions are
\begin{align}  \nonumber
D^0 \to \pi^+ \pi^- l^+l^- \, , \quad   &D^0 \to K^+ K^-l^+l^- \, , \\
  &D^+ \to K^+ \bar K^0 l^+ l^- \, ,  \label{eq:modes}\\
 D_s \to K^+ \pi^0 l^+l^- \, , \quad& D_s \to K^0 \pi^+ l^+ l^- \,  ,  \nonumber
 \end{align}
which all are  singly-Cabibbo suppressed.
We do not consider $D \to \pi^+ \pi^0ll $ decays  because isospin-conserving BSM contributions, such as those we are interested in this works, drop out in the  isospin limit.
However, this mode can complement SM tests in  hadronic 2-body decays of charm \cite{Bigi:2011em,Brod:2012ud,Hiller:2012xm,Muller:2015lua}.
Experimental results on four-body decays exist from LHCb  for branching ratios \cite{Aaij:2017iyr} of $D^0$ decays into muons and from
BESIII  for upper limits on branching ratios \cite{Ablikim:2018gro} of $D^0, D^+$ decays into electrons.

The aim of this work is to   study  the  angular distribution in $D \to P_1P _2 l^+ l^-$ decays on and off resonance, and to work out opportunities  for BSM signals.
Related distributions in $B \to K  \pi l^+ l^-$ decays have been analyzed in \cite{Das:2014sra}.
We describe non-resonant contributions with  an operator product expansion (OPE) in $1/Q$, $Q= \{\sqrt{q^2},m_c\}$, applicable at $q^2= {\cal{O}} (m_c^2)$ and detailed for $B \to V l^+ l^-$ decays in   \cite{Grinstein:2004vb}. Here,
$q^2$ denotes the dilepton invariant mass-squared and $m_c$ is the charm mass.
$D \to P_1 P_2$ form factors are available from
heavy hadron chiral perturbation theory (HH$\chi$PT)  \cite{Lee:1992ih}.
To capture the phenomenology we model resonance effects, which dominate the decay rates, assuming factorization and vector meson dominance, as in \cite{Cappiello:2012vg},
amended by  data \cite{Aaij:2017iyr}.

Despite the significant hadronic uncertainties there are  features in the SM  which are sufficiently clean  to warrant 
phenomenological exploitation of semileptonic rare charm decays:  negligible contributions to  axial-vector lepton coupling, $C_{10}^{(\prime)}$, and the suppression of 
CP, lepton flavor and lepton universality violation.
Our proposal to test the SM with $D \to P_1P _2 l^+ l^-$ decays is based on these features, which allow to perform  null tests  and to identify new physics. An interpretation in terms of BSM couplings, however, will again be subject to hadronic uncertainties.

This paper is organized as follows:
In section \ref{sec:weak} we review the weak Lagrangian, SM values and constraints on $|\Delta c| =|\Delta u|=1$ couplings.
The $D \to P_1 P_2  l^+ l^-$ angular distribution is given in  section \ref{sec:angular}.
Phenomenological resonance contributions are discussed in section \ref{sec:resonances}. 
BSM signals are worked out in section \ref{sec:BSM}, where we also discuss LNU-sensitive observables, probing BSM  interactions which distinguish between electrons and muons.
In section \ref{sec:con} we conclude. Auxiliary information on $D \to P_1 P_2  l^+ l^-$ matrix elements is given in the appendix.

\section{Weak Lagrangian \label{sec:weak}}

We consider BSM effects in the semileptonic operators,
\begin{align}
 &Q_9=(\bar u\gamma_{\mu}P_Lc)\left(\overline l\gamma^{\mu}l\right)\,, && Q_9'=(\bar u\gamma_{\mu}P_Rc)\left(\overline l\gamma^{\mu}l\right)\,,  \label{eq:Q9}\\
 &Q_{10}=(\bar u\gamma_{\mu}P_Lc)\left(\overline l\gamma^{\mu}\gamma_5l\right)  \, , &&  Q_{10}'=(\bar u\gamma_{\mu}P_Rc)\left(\overline l\gamma^{\mu}\gamma_5l\right)  \label{eq:Q10} \, ,  \\
 &Q_S=(\bar u P_R c)  ( \bar l l ),  &&              Q_S' =(\bar u P_Lc)  (\bar l l ) \,,  \label{eq:QS}\\
 &Q_P=(\bar u P_Rc) (\bar l \gamma_5 l)  \, , &&  Q_P'=(\bar u P_Lc) (\bar l \gamma_5 l)   \label{eq:QP} \,  , \\
  &Q_T=\frac{1}{2} (\bar u \sigma^{\mu \nu} c) (\bar l \sigma_{\mu \nu} l)  \, , &&  Q_{T5}'= \frac{1}{2} (\bar u \sigma^{\mu \nu} c) (\bar l  \sigma_{\mu \nu} \gamma_5 l)   \label{eq:QT} \,  ,
\end{align}
in the effective Lagrangian
\begin{align}
 \mathcal L_\text{eff}^\text{weak}=\frac{4G_F}{\sqrt 2}\frac{\alpha_e}{4\pi}\left(\sum_{q=d,s}V_{cq}^*V_{uq}\sum_{i=1}^2C_iQ_i^{(q)}
 +\sum_{i=9,10,S,P}   \left(C_iQ_i+C_i'Q_i'\right) + C_T Q_T +C_{T5} Q_{T5} \right)\,,
\end{align}
where $G_F$ is the Fermi constant, $\alpha_e$ denotes the fine structure constant and $V_{ij}$ are CKM matrix elements. $P_L, P_R$ denote left- and right-chiral projectors, respectively.

In the SM, the four-quark operators $Q_{1,2}^{(q)} \sim (\bar u \gamma_\mu P_L   q) ( \bar q \gamma^\mu P_L c)$ give rise to the dominant contributions to the branching ratios in  $|\Delta c| =|\Delta u|=1$ decays.
The Wilson coefficients  of the BSM-sensitive operators given in (\ref{eq:Q9})-(\ref{eq:QT}), on the other hand, are subject to an efficient GIM-cancellation, and suppressed.
At the charm mass scale  $\mu=m_c$ at NNLO  \cite{deBoer:2015boa,deBoer:2016dcg,deBoer:2017way},
\begin{align}
|C_{7}^{\rm eff}| \simeq\mathcal O(0.001) \, , \quad \quad   |C_{9}^{\rm eff}|_{\rm high\,q^2} \lesssim 0.01 \, ,  \quad \quad  C_{10, S,P,T,T5}^{\rm SM} =0 \, .  \label{eq:SM}
\end{align}
Here, the coefficient of the dipole operator $Q_7=\frac{m_c}{e}(\bar u \sigma_{\mu \nu} P_Rc)F^{\mu \nu}$, where $F^{\mu \nu}$ denotes the electromagnetic  field strength tensor, is also given for completeness.
The effective coefficients $C_{7,9}^{\rm eff}$ equal $C_{7,9}$ up to matrix elements of 4-quark operators which relax the GIM-cancellation, thus being the dominant contribution \cite{deBoer:2015boa,deBoer:2017way} and inducing a $q^2$-dependence, see \cite{deBoer}.

In addition, all primed coefficients $C^\prime_i$ are negligible in the SM.
Experimental constraints, available from the upper limit on the $D^+ \to \pi^+ \mu^+ \mu^-$ branching ratio,  and $D^0 \to \rho^0 \gamma$ are presently very weak, at least about two orders of magnitude away from the SM  \cite{deBoer:2015boa,deBoer:2017que}
\begin{align} \label{eq:mubounds}
  |C_7^{(\prime)}| \lesssim0.3 \, , \quad  \quad  |C_{9,10}^{(\prime)}| \lesssim 1  \, , \quad \quad  |C_{T,T5}| \lesssim 1 \, , \quad \quad  |C_{S,P}^{(\prime)}| \lesssim 0.1 \, ,
\end{align}
see  \cite{deBoer:2015boa} for correlated constraints.
Corresponding constraints on $c \to u e^+e^-$ processes are about  a factor  2-4 (5 times for $C_{T,T_5}$) weaker than the ones in (\ref{eq:mubounds}) on dimuons.
Constraints on  LFV processes $c \to u e^\pm \mu^\mp$  are 6-7 times (4 times for $C^{(\prime)}_{S,P}$) weaker than the dimuon constraints.
To discuss LNU or LFV, Wilson coefficients and operators become lepton-flavor dependent. To avoid clutter,  we refrain from showing 
 lepton flavor superscripts throughout this paper.

\section{Full angular distribution \label{sec:angular}}

In section \ref{sec:general} we discuss the  full angular distribution for $D \to P_1P_2  l^+ l^-$ decays and identify SM null tests that exist thanks  to the extreme GIM-suppression in charm.
In section \ref{sec:OPE} we give the angular distribution in the  low  hadronic recoil OPE, which defines a factorization-type framework at leading order in $1/m_c$.
To estimate possible BSM signals, which involve SM-BSM interference, we need to estimate  SM contributions to decay amplitudes as well.
The phenomenological description of  the dominant resonance-induced contributions  is detailed in section \ref{sec:resonances}.

\subsection{General case \label{sec:general}}

 The $D \to P_1 P_2  l^+ l^-$ angular distribution, with the angles $\theta_l, \theta_{P_1}, \phi$ defined as in \cite{Bobeth:2008ij} taking into account  footnote 2 of Ref.~\cite{Bobeth:2012vn},
can be written as 
  \begin{eqnarray}   \label{eq:full}
d^5\Gamma 
&=&\frac{1}{ 2  \pi} \left[ \sum c_i(\theta_l,\phi) I_i (q^2,p^2,\cos \theta_{P_1}) \right] dq^2dp^2d\cos\theta_{P_1}d\cos\theta_l d\phi\,,
\end{eqnarray}
where  $q^2$, $p^2$ denotes the invariant mass-squared of the dileptons, ($P_1 P_2$)-subsystem, respectively, and 
\begin{align}
c_1 & =1\,, \quad c_2=\cos 2\theta_l\,, \quad c_3=\sin^2\theta_l\cos 2\phi\,, \quad c_4=\sin 2\theta_l \cos \phi\,, \quad c_5=\sin\theta_l\cos\phi\,, \nonumber \\ c_6& =\cos\theta_l\,, \quad c_7=\sin\theta_l\sin\phi\,, \quad c_8=\sin 2\theta_l\sin\phi\,, \quad c_9=\sin^2\theta_l\sin2\phi \, .
\label{eq:ci}
\end{align}
$\theta_l$ denotes the angle  between the $l^-$-momentum and the $D$-momentum in the dilepton center-of-mass  system (cms), $\theta_{P_1} $ is the angle  between the $P_1$-momentum and the negative direction of flight of the $D$-meson in the ($P_1P_2$)-cms, and $\phi$ is the angle between the normals of the ($P_1P_2$)-plane and the ($ll$)-plane in the $D$ rest frame.
The angles are within the ranges
\begin{align}
-1 < \cos \theta_{P_1} \leq 1 \, , \quad -1 < \cos \theta_l  \leq 1 \, , \quad  0< \phi  \leq 2 \pi \, .
\end{align}
$P_1$ is the meson that contains the quark emitted from the semileptonic weak $\bar u c ll$ vertex.
For instance, $P_1=\pi^+$ and $P_1=K^+$ in the $ D^0, D^+$-decays in (\ref{eq:modes}).

The angular coefficients $I_i \equiv I_i(q^2,p^2,\cos \theta_{P_1})$  are given in terms of transversity amplitudes~\footnote{No tensor and  no (pseudo)-scalar operators included, and for vanishing lepton mass.} as 
\begin{align}  \nonumber
I_1 & = \phantom{-}\frac{1}{16} \bigg[ |\azeL|^2 +(L\to R) + \frac{3}{2}\sin^2 \theta_{P_1} \{ |\apeL|^2 + |\apaL|^2 + (L\to R) \} \bigg]\,,
\\  \nonumber
  I_2 & = -\frac{1}{16} \bigg[ |\azeL|^2 + (L\to R)  -\frac{1}{2} \sin^2 \theta_{P_1} \{ |\apeL|^2+ |\apaL|^2 + (L\to R) \}\bigg]\,,
\\  \nonumber
  I_3 &=  \phantom{-}\frac{1}{16}   \bigg[ |\apeL|^2 - |\apaL|^2  + (L\to R)\bigg]  \sin^2\theta_{P_1}\,,
\\  \nonumber
  I_4 & = -\frac{1}{8}  \bigg[\re (\azeL^{}\apaL^*) + (L\to R)\bigg] \sin\theta_{P_1}\,,
\\  \label{eq:IH}
  I_5 & = -\frac{1}{4}  \bigg[\re(\azeL^{}\apeL^*) - (L\to R)\bigg] \sin\theta_{P_1}\,,
\\  \nonumber
  I_6 & = \phantom{-}\frac{1}{4}   \bigg[\re (\apaL^{}\apeL^*) - (L\to R)\bigg] \sin^2\theta_{P_1}\,,
\\  \nonumber
  I_7 &=- \frac{1}{4} \bigg[\im (\azeL^{}\apaL^*) - (L\to R)\bigg] \sin\theta_{P_1}\,,
\\  \nonumber
  I_8 & =-\frac{1}{8}  \bigg[\im(\azeL^{}\apeL^*) + (L\to R)\bigg] \sin\theta_{P_1}\,,
\\  \nonumber
  I_9 & =\phantom{-}\frac{1}{8}   \bigg[\im (\apaL^{*} \apeL) + (L\to R)\bigg] \sin^2\theta_{P_1}\,.
\end{align}
The subscript $0, \parallel$ and $\perp$ stands for longitudinal, parallel and perpendicular polarization, respectively.
Here, $L,R$ denotes the handedness of the lepton current. In the SM electromagnetically-induced contributions dominate $c \to  u l^+l^-$ transitions due to the GIM-mechanism (\ref{eq:SM}). Hence, by inspecting the relative signs between the left-handed and the right-handed contributions in (\ref{eq:IH}), it follows that $I_{5,6,7}$ constitute null tests, as they require axial-vector contributions to be non-vanishing.

One may wonder about backgrounds to $I^{\rm SM}_{5,6,7}=0$.
Intermediate pseudo-scalar resonances $D \to  P_1 P_2 \eta^* \to P_1P_2  l^+ l^-$ induce a  contribution to pseudo-scalar operators
$Q_P$ not included in (\ref{eq:IH}). The impact can be read off  from the $D \to V( \to P_1P_2) l^+ l^-$ angular distribution \cite{Bobeth:2012vn}:
Contributions from $C_P$ to $I_{5,6,7}$ require the presence of tensor operators.
Similarly, lepton mass effects pose  no challenge to the null tests, as finite $m_l$ contributions require the presence of scalar or tensor operators which are both negligible in the SM (\ref{eq:SM}).
Finite SM contributions to axial-vector couplings are expected to arise from higher order  electromagnetic effects.  For instance, a 2-loop diagram with  an insertion of
$Q_{1,2}^{(q)}$ with two photons induces a contribution at the relative order  $\alpha_e/(4 \pi)$, about permille level.
We estimate contributions from electromagnetic operator mixing 
as  $C_{10}<0.01\,C_9$ \cite{Bobeth:2003at,Huber:2005ig,deBoer}, which is small,  at most $10^{-4}$ in the SM (\ref{eq:SM}).
As will be shown in section \ref{sec:I567}, order one BSM contributions are needed to generate finite angular coefficients up to few percent.
Therefore, higher order effects are of no concern to the null tests $I_{5,6,7}$ within the accuracy that can be achieved in the  foreseeable future, $3 \% (1 \%)$  at Run II (upgrade) on 
$D^0 \to \pi^+ \pi^- \mu^+ \mu^-$ asymmetries at LHCb \cite{Vacca:2015ffa}.
We learn that angular analysis in charm is  simpler than in $B$-decays {\it because} charm is dominated by resonances.

Integrating (\ref{eq:full}) over $\phi, \cos \theta_l$ and both, respectively, yields the decay distributions
  \begin{eqnarray}
\frac{d^4\Gamma }{dq^2dp^2d\cos\theta_{P_1}d\cos\theta_l}
&=&  I_1  +I_2 \cos  2\theta_l  +I_6  \cos \theta_l   \,, \\
\frac{d^4\Gamma }{dq^2dp^2d\cos\theta_{P_1}d \phi}
&=&\frac{1 }{ \pi } \left(   I_1  - \frac{I_2}{3}   + \frac{\pi}{4} I_5  \cos \phi + \frac{\pi}{4}  I_7 \sin \phi + \frac{2}{3} I_3 \cos 2 \phi +  \frac{2}{3} I_9 \sin 2 \phi  \right) \,, \\
\frac{d^3\Gamma }{dq^2dp^2d\cos \theta_{P_1} }
&=&2  \left(   I_1  - \frac{I_2}{3}  \right) \,. \label{eq:dq2p2cth}
\end{eqnarray}
The forward-backward asymmetry in the leptons, $A_{\rm FB} \propto I_6$ can be obtained  from asymmetric $\cos\theta_l$ integration
\begin{align} \label{eq:i6}
I_6 = \frac{1}{2} \left[ \int_0^1 d\cos\theta_l  - \int_{-1}^0  d\cos\theta_l \right] \frac{d^4\Gamma }{dq^2dp^2d\cos\theta_{P_1}d\cos\theta_l} \, . 
\end{align}
The observables $I_7$ and $I_5$ can be obtained, for instance, as follows
\begin{align}
I_7  & = \ \left[ \int_0^{\pi}  d  \phi  - \int_{\pi}^{2 \pi}  d \phi  \right]  \frac{d^4\Gamma }{dq^2dp^2d\cos\theta_{P_1}d \phi} \, , \\
I_5  & = \ \left[ \int_{-\pi/2}^{\pi/2}  d  \phi  - \int_{\pi/2}^{3 \pi/2
}  d \phi  \right]  \frac{d^4\Gamma }{dq^2dp^2d\cos\theta_{P_1}d \phi}   \, . 
\label{eq:i5}
\end{align}
Methods to get angular coefficients for P-wave contributions are given in \cite{Bobeth:2008ij}.

At the kinematic end point of zero hadronic recoil   the following exact relations hold  \cite{Das:2014sra}
\begin{align} \label{eq:ep}
I_3=-\frac{I_1+I_2}{2}\, , \quad
I_4 = - \sqrt{ \frac{( I_1+I_2)(I_1-3 I_2)}{2} } \, , \quad
I_{5,6,7,8,9} =0 \, .
\end{align}

The corresponding observables of the CP-conjugated  $\bar D$ decays are given by $I_{1,2,3,4,7}\to\bar I_{1,2,3,4,7}$ and $I_{5,6,8,9}\to-\bar I_{5,6,8,9}$, where $\bar I$ equals $I$ with the  weak phases
flipped.  In $\bar D$-decays,  $\theta_l$ is the angle between the $l^-$-momentum and $\bar D$-momentum in the dilepton cms, $\theta_{P_1}$ is the angle between the $P_1$-momentum and the negative $\bar D$-momentum in the  ($P_1P_2$)-cms, and $\phi$ the angle between the ($P_1P_2$)- and ($ll$)-planes.  We keep the definition of $P_1$ from $D$ decays  for $\bar D$  decays.

The observables $I_{7,8,9}$  are 
(naive) T-odd and corresponding CP asymmetries are not suppressed by small strong phases.
The  observables $I_{5,6,8,9}$ are odd under the CP-transformation. Therefore, if distributions from (untagged) $D^0$ and $\bar D^0$  decays are averaged one measures a CP-asymmetry, $A_k, k=5,6,8,9$.
Due to the smallness of $V_{cb}^* V_{ub}/(   V_{cs}^* V_{us}   )$ these constitute null tests of the SM.
Note that time-dependent effects in angular observables \cite{Bobeth:2008ij}  are suppressed by the  small  $D^0-\bar D^0$ width difference \cite{Amhis:2016xyh}.

\subsection{OPE and factorization  \label{sec:OPE}}

At leading order low recoil OPE,  long- and short-distance physics factorizes
as follows  \cite{Das:2014sra}
\begin{align} \nonumber
I_1 & = \phantom{-}\frac{1}{8} \left[   |{\cal F}_0 |^2 \rho_1^- +\frac{3}{2} \sin^2 \theta_{P_1}  \{  |{\cal F}_\parallel |^2 \rho_1^-+  |{\cal F}_\perp |^2 \rho_1^+\} \right]  \, ,  \\ \nonumber
I_2 & =- \frac{1}{8} \left[   |{\cal F}_0 |^2 \rho_1^- -\frac{1}{2} \sin^2 \theta_{P_1}  \{  |{\cal F}_\parallel |^2 \rho_1^-+  |{\cal F}_\perp |^2 \rho_1^+\} \right]   \, ,  \\ \nonumber
I_3 & = \phantom{-}\frac{1}{8}    \left[ |{\cal F}_\perp |^2 \rho_1^+ -  |{\cal F}_\parallel |^2 \rho_1^-  \right]  \sin^2 \theta_{P_1}  \, ,  \\ \nonumber
I_4 &= - \frac{1}{4} {\rm Re}({\cal F}_0 {\cal F}_\parallel^*) \,\rho_1^-  \sin \theta_{P_1}  \, ,  \\
\label{eq:Iope}
I_5 &=  \phantom{-}\left[{\rm Re}({\cal F}_0 {\cal F}_\perp^*)  {\rm Re} \rho_2^++ {\rm Im} ({\cal F}_0 {\cal F}_\perp^*) 
{\rm Im} \rho_2^-  \right] \sin \theta_{P_1}  \, ,  \\ \nonumber
I_6&= - \left[{\rm Re}({\cal F}_\parallel {\cal F}_\perp^*) {\rm Re} \rho_2^+ + {\rm Im} ({\cal F}_\parallel {\cal F}_\perp^*) {\rm Im} \rho_2^-  \right]\sin^2 \theta_{P_1}  \, ,   \\ \nonumber
I_7&=   {\rm Im} ({\cal F}_0 {\cal F}_\parallel^*)\,\delta \rho\,  \sin \theta_{P_1} \, ,   \\ \nonumber
I_8&=  \frac{1}{2} \left[ {\rm Re}({\cal F}_0 {\cal F}_\perp^*) {\rm Im} \rho_2^+  -  {\rm Im} ({\cal F}_0 {\cal F}_\perp^*) {\rm Re} \rho_2^-  \right] \sin \theta_{P_1} \, ,   \\ \nonumber
I_9&=  \frac{1}{2} \left[{\rm Re}({\cal F}_\perp {\cal F}_\parallel^*) {\rm Im} \rho_2^+ + {\rm Im} ({\cal F}_\perp {\cal F}_\parallel^*) {\rm Re} \rho_2^- \right]\sin^2 \theta_{P_1}   \, ,  \nonumber
\end{align}
where the short-distance coefficients read
\begin{align} \nonumber
\rho_1^\pm & = \left| C_9^{\rm eff} \pm  C_9^\prime \right|^2 + |C_{10}  \pm  C_{10}^\prime|^2 \, , \\ \nonumber
\delta \rho & = {\rm Re}\left[ \left(C_9^{\rm eff} -  C_9^\prime  \right)\left(C_{10}  -  C_{10}^\prime\right)^* \right]  \, , \\ \nonumber
{\rm Re} \rho_2^+ & ={\rm Re} \left[  C_9^{\rm eff}   C_{10}^*  - C_9^\prime  C_{10}^{\prime *} \right]  \, , \\
{\rm Im} \rho_2^+ & = {\rm Im}\left[     C_{10}^{\prime} C_{10}^* 
+ C_{9}^{\prime}       C_9^{{\rm eff}   *}
  \right]  \, ,       \label{eq:BSM-dep} \\ \nonumber
{\rm Re} \rho_2^- & = \frac{1}{2} \left[  |C_{10}|^2  -| C_{10}^\prime|^2+ \left|C_{9}^{\rm eff} \right|^2  -\left| C_{9}^\prime \right|^2 
\right]  
\, , \\ \nonumber
{\rm Im} \rho_2^- & = {\rm Im} \left[C_{10}^\prime  C_9^{{\rm eff}  *}- C_{10} C_9^{\prime *} \right]  \, .
\end{align}
As we are anticipating BSM contributions to semileptonic operators  (\ref{eq:Q9}), (\ref{eq:Q10}) only~\footnote{BSM effects in dipole operators can be tested in radiative $D$-decays, {\it e.g.},    \cite{deBoer:2018zhz,Isidori:2012yx,deBoer:2017que}.}
we dropped the contributions from dipole  operators, which  enter as $\propto ( m_c m_D/q^2)  C_7^{\rm eff}$ for clarity.
Full formulae can be seen in  \cite{Das:2014sra}. We explicitly checked that contributions from dipole  operators are  negligible for the purpose of our analysis.

The  transversity form factors
 ${\cal{F}}_i$, $i=0, \perp, \parallel$ can be written as
\begin{align} \nonumber 
{\cal{F}}_0 &= \frac{{\cal N}_{\rm nr}}{2}  \bigg[  \lambda^{1/2 }w_+(q^2,p^2,\cos \theta_{P_1})+\frac{1}{p^2}\{(m_{P_1}^2  -m_{P_2}^2)\lambda^{1/2} \\ \nonumber
& \quad \quad \quad   -(m_D^2-q^2-p^2) \lambda^{1/2}_{p}\cos \theta_{P_1}\} w_-(q^2,p^2,\cos \theta_{P_1}) \bigg]\,,\\
\label{eq:Fi}
{\cal{F}}_\parallel &= {\cal N}_{\rm nr}  \sqrt{ \lambda_p \frac{q^2}{p^2}} \, w_-(q^2,p^2,\cos \theta_{P_1})\,, \qquad
{\cal{F}}_\perp = \frac{{\cal N}_{\rm nr}}{2}\sqrt{ \lambda \lambda_p \frac{  q^2}{p^2}} \, h(q^2,p^2,\cos \theta_{P_1})\,, \\
 & \quad \quad  \quad \quad   {\cal{N}}_\text{nr}=\frac{G_F\alpha_e}{2^7\pi^4m_D}\sqrt{\pi\frac{\sqrt{  \lambda \lambda_p }}{m_Dp^2}} \, .   \nonumber
\end{align} 
Here, $\lambda=\lambda(m_D^2,q^2,p^2)$ and $\lambda_p=\lambda(p^2,m_{P_1}^2,m_{P_2}^2)$, where $\lambda(a,b,c)=a^2+b^2+c^2-2 (ab + ac+ bc)$.
The  $D \to P_1 P_2$ transition form factors  are defined as
\begin{align}
\langle P_1(p_1)P_2(p_2)|\bar u\gamma_\mu(1-\gamma_5)c|
D(p_D)\rangle& 
		=i \left[w_+p_\mu+w_-P_\mu+rq_\mu+ih\epsilon_{\mu\alpha\beta\gamma}p_D^\alpha p^\beta P^\gamma\right]\,, \label{eq::FFLLW} \\
       \langle P_1(p_1)P_2(p_2)|
\bar u i q^\nu \sigma_{\mu \nu}(1+\gamma_5)   c|D(p_D)\rangle &=
- i m_D \! \left[ w_+' p_{ \mu} + w_-' P_{\mu} + r' q_{\mu} \!
       + i h' \varepsilon_{\mu\alpha\beta\gamma}p_D^\alpha p^\beta
       P^\gamma \right]  \, , \label{eq::FFLLW2}
\end{align}
where the right-hand sides  have to be multiplied by an isospin factor of $1/\sqrt{2}$  for every neutral pion in the final state and  we tacitly suppressed the dependence on $q^2,p^2$ and $\cos \theta_{P_1}$ in the form factors. Here, $q^\mu=p_+^\mu+p_-^\mu$, $p^\mu=p_1^\mu+p_2^\mu=p_D^\mu-q^\mu$ and $P^\mu=p_1^\mu-p_2^\mu$.
Since the dipole operators in the SM are negligible and we do not consider BSM tensor operators,  the dipole form factors  (\ref{eq::FFLLW2}) are not needed for  our analysis. 
$r^{(\prime)}$ does not contribute to $D \to P_1 P_2 l^+ l^-$ decays for $m_l=0$~\footnote{(Pseudo)-scalar operators would also require $r$,
$\langle P_1(p_1)P_2(p_2)|\bar u (1+\gamma_5)c|
D(p_D)\rangle =i /m_c\left[ w_+ p \cdot q + w_- P \cdot q + r q^2 \right]$. 
}.

The relevant non-resonant $D \to P_1 P_2$ form factors  $w_\pm,h$  are available from HH$\chi$PT \cite{Lee:1992ih}. Numerical input is  given in the appendix.
Note that 
HH$\chi$PT applies if the participating light mesons are sufficiently soft.  We find that $E_\pi -m_\pi$ in the $D$-meson's cms in $D\to \pi^+ \pi^-  l^+ l^-$ decays does not exceed $0.4$ (0.6) GeV for $q^2$ above $m_\phi^2$ ($m_\rho^2$), where $E_\pi$ denotes the energy of any of the  pions  in the $D$-cms.
The region  above $m_\phi^2$  is kinematically closed for  $D\to K^+ K^-  l^+ l^-$ decays. In these decays  $E_K -m_K$ in the $D$-cms 
does not exceed $0.3$ GeV for  all $q^2$, where $E_K$ denotes the energy of any of the  kaons in the $D$-cms.
Although formally they are limited to low hadronic recoil we use the HH$\chi$PT form factors  in the full phase space also for  $D\to \pi^+ \pi^-  l^+ l^-$ in absence of other estimates.

\begin{figure}
\begin{center}
\includegraphics[width=0.45\textwidth]{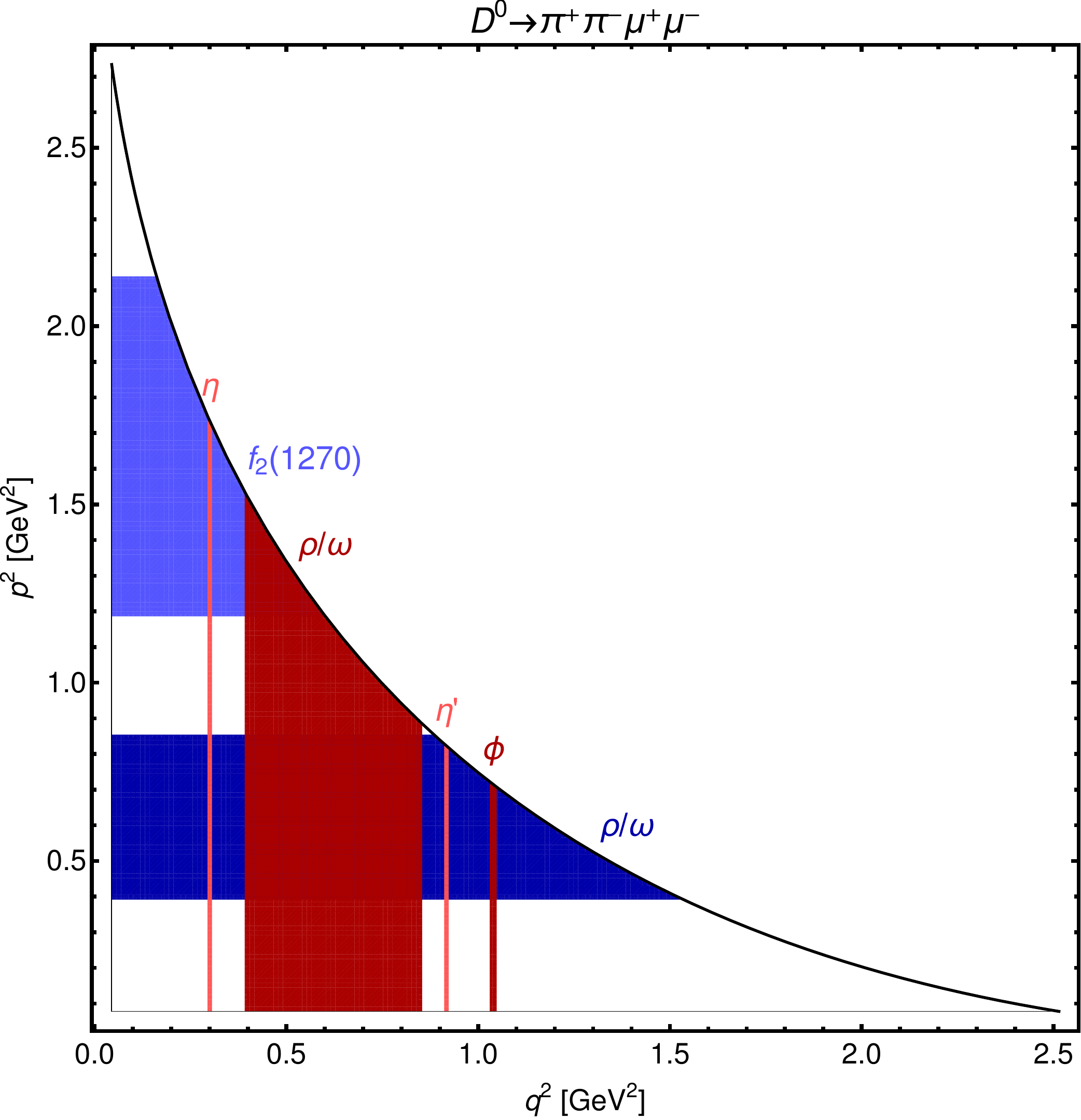}\qquad
\includegraphics[width=0.45\textwidth]{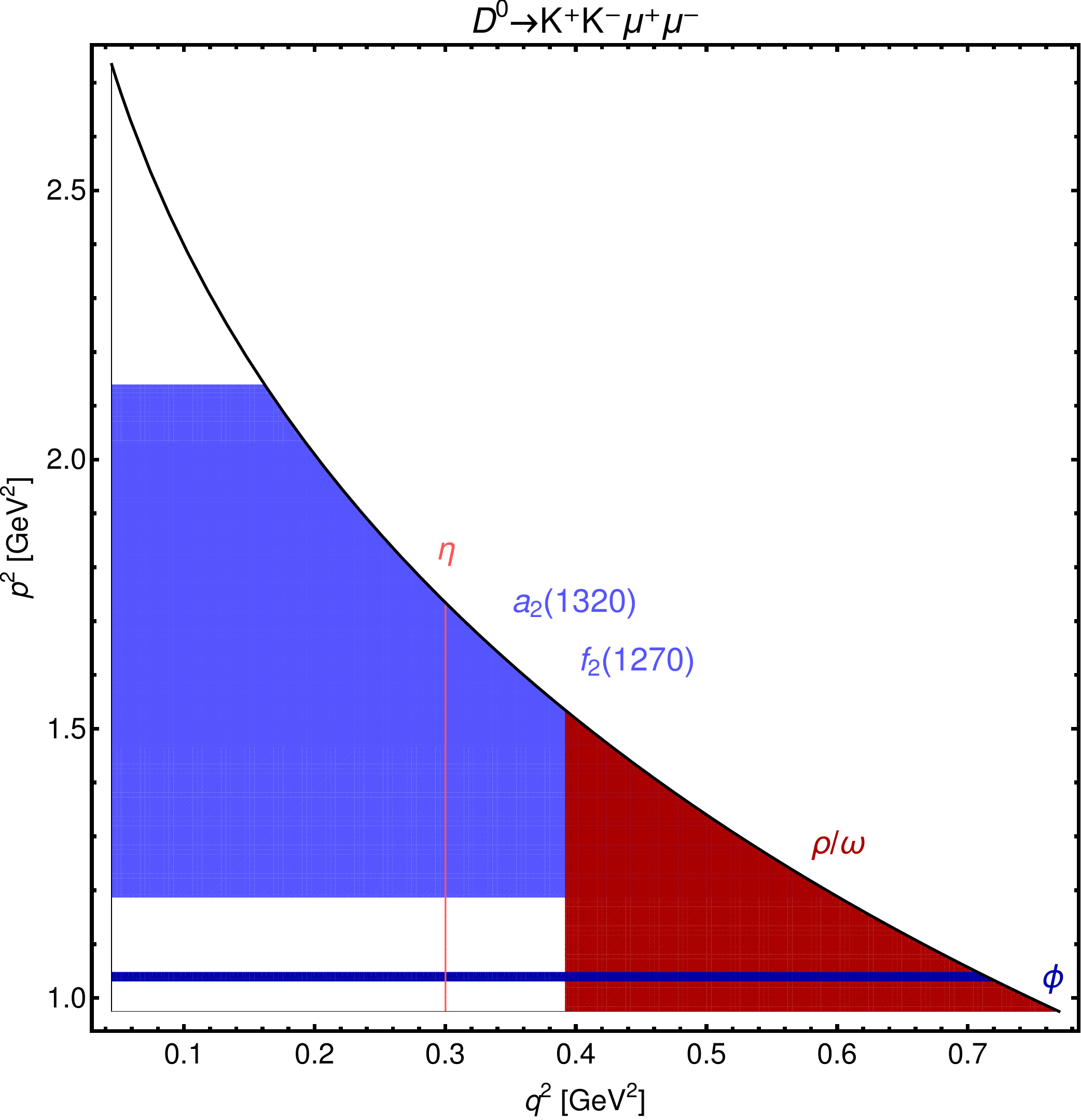}
\end{center}
\caption{Phase space and dominant  resonances in $q^2$ and $p^2$ for $D^0 \to \pi^+ \pi^- \mu^+ \mu^-$  decays (left) and $D^0 \to K^+ K^- \mu^+ \mu^-$  decays (right).
The bands correspond to $(\rm mass\pm width)^2$.
The very wide scalar resonances  $f_0(500)$ and $f_0(980)$ would fill everything below the $f_2$ in the $\pi \pi$ plot and are not shown.
}
\label{fig:q2p2}
\end{figure}

We use this prescription, factorization  plus HH$\chi$PT form factors, for the BSM short-distance contributions from 4-fermion operators (\ref{eq:Q9}), (\ref{eq:Q10}) to estimate BSM signals in the whole phase space for both $\pi \pi $ and $KK$ modes. 
In  figure  \ref{fig:q2p2} the $q^2,p^2$-phase space for $D^0\to\pi^+\pi^- l^+ l^-$ (plot to the left) and $D^0\to K^+K^- l^+ l^-$ decays  (plot to the right) is shown with
dominant resonances.
The OPE formally applies for  $q^2={\cal{O}}(m_c^2)$. This is approximately the region above the $\phi$-peak in $D \to \pi^+ \pi^- l^+ l^-$ decays, and nowhere in $D \to K^+ K^-  l^+ l^-$.
QCDF at least formally  works for  $p^2={\cal{O}}(\Lambda^2)$ and  $p^2 \sim q^2$, that is, when the $(P_1P_2)$-system is light and energetic in the $D$-cms, see also \cite{Grinstein:2005ud}.
While QCDF therefore can be used in $D^0\to\pi^+\pi^- l^+ l^-$ for low $q^2$, this  region is mostly occupied by resonances. In $D \to K^+ K^-  l^+ l^-$  with $p^2_{\rm min} \approx 1 \, \mbox{GeV}^2$ 
there is little room left.
For  $p^2={\cal{O}}(m_c^2)$  the dilepton system is soft  in the $D$-cms.
A related discussion of phase space has been given in \cite{Faller:2013dwa} for  $B \to \pi \pi l \nu$ decays. 
Due to the lower value of the heavy quark mass the phase space in charm is much more compressed  than in $b$-decays.

\section{Resonance contributions  \label{sec:resonances}}

Several resonances contribute to $D \to P_1P_2 l^+ l^-$ decays.
First we consider resonances  in the ($P_1 P_2$)-subsystem, that is, 
in  $p^2$. Depending on the spin $j=0,1,...$ of  the resonance,  such contributions are termed S,P,...-wave, respectively.
Due to the lower mass of the $D$-mesons relative to the $B$-ones, there are fewer resonances and ones with lower spin contributing in charm.
Lowest lying resonances with sizable branching ratios into $ \pi \pi$ are the $\rho$ and scalars $\sigma=f_0(500)$ and $f_0(980)$.
At spin 2 there is the  $f_2(1270)$.
For $K^+K^-$, it is essentially the $\phi$,  and 
for $K \pi$  there is the $K^*(892)$, the scalars $\kappa$ and $K_0^*(1430)$ and the spin 2 resonance $K_2^*(1430)$.

 We model the resonance structure  in $p^2$ for $D^0 \to \pi^+ \pi^- l^+ l^-$ decays  by the $\rho$-contribution, which is dominant at least in the wider vicinity of  $p^2 \approx m_\rho^2$.
  $D \to \rho$ form factors are taken from \cite{Melikhov:2000yu}, see appendix.
  D-waves and higher are phase space suppressed  relative to the $\rho$ and contribute to small $q^2 \lesssim 0.4 \,  \mbox{GeV}^2$ only.
  Further study including scalar contributions, which are rather wide and less known,
is beyond the scope of this work, which aims at identifying  null tests and illustrating the  sensitivity to BSM physics. We stress, however,  that  since there  is no S-wave contribution to 
$I_{3,6,9}$~\cite{Das:2014sra} these angular coefficients are unaffected by scalars.  In addition, the S-P  interference terms in  $I_{4,5,7,8}$ can be separated from the P-wave contribution by angular analysis,
therefore  scalars can be experimentally subtracted in these coefficients.

The other type of resonances contribute in  $q^2$ as  $D \to P_1P_2  \gamma^*$, $\gamma^* \to l^+ l^-$ via $\omega, \rho^0, \phi$ and $\eta^{(\prime)}$.
We model these  contributions  with a  phenomenological Breit-Wigner shape for $C_9 \to C_9^\text R$ for vector 
and $C_P \to C_P^\text R$ for pseudoscalar mesons  \cite{deBoer:2015boa,Fajfer:2005ke}
\begin{align}
 C_9^\text R&=a_\rho e^{i\delta_\rho} \left(\frac1{q^2-m_\rho^2+im_\rho\Gamma_\rho}-\frac13\frac1{q^2-m_\omega^2+im_\omega\Gamma_\omega}\right)+\frac{a_\phi e^{i\delta_\phi}}{q^2-m_\phi^2+im_\phi\Gamma_\phi}  \, , \nonumber\\
 C_P^\text R &=\frac{a_\eta e^{i\delta_\eta}}{q^2-m_\eta^2+im_\eta\Gamma_\eta}+\frac{a_{\eta'}}{q^2-m_{\eta'}^2+im_{\eta'}\Gamma_{\eta'}}\, \, ,
 \label{eq:BW}
\end{align}
where $m_M,\Gamma_M$ denotes the mass and total width, respectively, of the resonance $M=\eta^{(\prime)}, \rho^0, \omega, \phi$, and we used isospin to relate the $\rho^0$ to the $\omega$.
Corresponding  transversity form factors  are given in the appendix, eqs.~(\ref{eq:fullformfactor})-(\ref{eq:Fperp}). 
LHCb  \cite{Aaij:2017iyr} has provided  branching ratios in $q^2$-bins  around the resonances $\rho/\omega$ and $\phi$, 
\begin{align} \label{eq:rhobin}
\mathcal B (D^0\to\pi^+\pi^-\mu^+\mu^-)|_{[0.565-0.950]\,\text{GeV}}& =(40.6\pm5.7)\times 10^{-8} \, , \\    \label{eq:phibin}
 \mathcal B (D^0\to\pi^+\pi^-\mu^+\mu^-)|_{[0.950-1.100]\,\text{GeV}}&=(45.4\pm5.9)\times 10^{-8} \, , \\
\mathcal  B(D^0\to K^+K^-\mu^+\mu^-)|_{[>0.565]\,\text{GeV}}&=(12.0\pm2.7)\times 10^{-8} \, ,  \label{eq:rhobinKK}
\end{align} 
where we added uncertainties in quadrature and neglected correlations.
The resonance parameters in  $C_{9,P}^{\text R}$ are in general $p^2$-dependent. We assume that the dominant $p^2$-dependence is taken care of by the $\rho$-lineshape specified in the appendix  such that
the $a_M$ are fixed by (\ref{eq:rhobin}), (\ref{eq:phibin}) at $p^2 \approx m_\rho^2$:
\begin{align}\label{eq:resonance_parameters}
 a_\phi^{\pi\pi}\simeq0.3\,\text{GeV}^2\,,\quad a_\rho^{\pi\pi}\simeq0.7\,\text{GeV}^2\,.
\end{align} 
For $M=\eta^{(\prime)}$ we use ${\cal{B}}(D^0 \to \pi^+  \pi^- M (\to \mu^+ \mu^- )) \simeq  {\cal{B}}(D^0 \to M \pi^+\pi^-) {\cal{B}}(M \to \mu^+ \mu^-)$ and take the right-hand side from data 
\cite{Patrignani:2016xqp} together with  ${\cal{B}}(\eta^\prime \to \mu^+ \mu^-) \sim {\cal{O}}(10^{-7})$ \cite{Patrignani:2016xqp,Landsberg:1986fd}. We obtain
\begin{align}\label{eq:C9R_parameters}
 a_\eta^{\pi\pi}\simeq0.001\,\text{GeV}^2\,,\quad a_{\eta'}^{\pi\pi}\sim0.001\,\text{GeV}^2\,.
\end{align}
To implement the pseudo-scalar contributions we employed the  $D \to \rho l^+ l^-$ distributions that can be inferred  from  \cite{Bobeth:2012vn}.
We note that fitting $M=\rho^0,\omega,\phi$ in the zero-width approximation \cite{Patrignani:2016xqp} and  using $2\times {\cal{B}}(D^0 \to \rho^0 \rho^0) \times {\cal{B}}(\rho^0 \to \mu^+ \mu^-)$ for the $\rho^0$,  one obtains parameters consistent with (\ref{eq:resonance_parameters}), with $a_\omega$ somewhat below the isospin prediction $a_\rho/3$ as already noticed for $D^+\to\pi^+\mu^+\mu^-$ \cite{deBoer:2015boa}.
The strong phases $\delta_M$ remain undetermined by this and introduce theoretical uncertainties.

The situation in the $D^0 \to K^+ K^- l^+ l^-$ channel is different  as the obvious resonance, the $\phi$, is not produced through a significant form-factor type contribution in $D^0$-decays.
The  small $u \bar u$ admixture in the $\phi$ should give approximately  few percent  of the corresponding $\rho \to \pi \pi$  amplitude. Similarly,
lowest lying mesons with larger $u \bar u$ content, $f_2(1270)$, $a_2(1320)$, decay with about 5\% branching ratio to $K \bar K$, which again is a correction.
The dominant contribution is expected to originate from 
annihilation topologies $D^0 \to \phi (\to K^+ K^-) \gamma^*$, recently  discussed in \cite{Feldmann:2017izn} for $D \to \rho l^+l^-$ decays within QCDF.
Here we continue following a phenomenological approach, as in \cite{Cappiello:2012vg}, based on factorization and  vector meson dominance, and use
   \begin{align}
 \langle \gamma^* (q) \phi(p) | C_1 Q^{(s)}_1 +C_2 Q^{(s)}_2  |D^0 (p_D)  \rangle \sim C_9^{\rm R} |_{a_\phi=0}  \cdot \langle V(q)  | \bar u \gamma^\mu P_L c | D^0 (p_D) \rangle \langle \phi(p) | \bar s \gamma_\mu s | 0 \rangle  \, ,
 \end{align}
where $ V=\rho^0,\omega$,  and we neglect differences between the $D \to \rho^0$ and $D \to \omega$ form factors.
The corresponding amplitude in $D \to \pi^+ \pi^- l^+ l^-$ decays, that is, when the $\rho^0$ which decays to $\pi^+ \pi^-$ is created at the weak vertex rather than through a form factor, is  effectively included  in our prescription with resonance parameters fixed by  data -- allowing for the extra amplitude would merely result in  re-fitting  $a_\phi^{\pi \pi}$ and $a_\rho^{\pi \pi}$~\footnote{There is a subtlety here, because the two contributions have slightly different $p^2$-behavior from the form factors. Since these are slowly varying functions, as opposed to the Breit-Wigner resonance shapes, this is a negligible  effect within the uncertainties and the purpose of this work.}.
Specifically,  for $D^0 \to K^+ K^- l^+ l^-$ decays
we  use $C_9^{\rm R}$ as in (\ref{eq:BW}) with $a_\phi=0$, and the transversity form factors  $F_{i \phi}$ given in the appendix.
The $\phi$-lineshape is parameterized by a Breit-Wigner distribution.
To include the contribution from  $\eta \to l^+ l^-$  we use 
   \begin{align}
 \langle \gamma^* (q) \phi(p) | C_1 Q^{(s)}_1 +C_2 Q^{(s)}_2  |D^0 (p_D)  \rangle \sim C_P^{\rm R} |_{a_{\eta^\prime}=0}  \cdot \langle \eta(q)  | \bar u \gamma^\mu P_L c | D^0 (p_D) \rangle \langle \phi(p) | \bar s \gamma_\mu s | 0 \rangle  \, .
 \end{align}
Note, the $\eta^\prime$ is kinematically forbidden.
We then obtain from (\ref{eq:rhobinKK}) and the zero-width approximation for the $\eta$ \cite{Patrignani:2016xqp}
\begin{align}
a_\rho^{KK}   \simeq 0.5 \,\text{GeV}^2\, \, , \quad    a_\eta^{KK}\simeq0.0003 \,\text{GeV}^2\,.
\end{align}

\begin{table}
 \centering
 \begin{tabular}{ccccc}
  \toprule
  branching ratio                              &  $D^0\to\pi^+\pi^-\mu^+\mu^-$  &  $D^0\to K^+K^-\mu^+\mu^-$     &  $D^0\to\pi^+\pi^-e^+e^-$  &  $D^0\to K^+K^-e^+e^-$  \\
  \midrule
  LHCb \cite{Aaij:2017iyr}$^\dagger$  &  $(9.64\pm1.20)\times10^{-7}$  &  $(1.54\pm0.33)\times10^{-7}$  &  --                        &  --  \\
  BESIII \cite{Ablikim:2018gro}       &  --                            &  --                            &  $<0.7\times10^{-5}$       &  $<1.1\times10^{-5}$  \\
  \midrule
  resonant                          &  $\sim1\times10^{-6}$          &   $\sim1\times10^{-7}$    &  $\sim10^{-6}$             &  $\sim10^{-7}$  \\
  non-resonant                        &  $10^{-10}-10^{-9}$          &   $\mathcal O(10^{-10})$       &  $10^{-10}-10^{-9}$             &  $\mathcal O(10^{-10})$  \\
   \cite{Cappiello:2012vg}         &  $\sim  10^{-6}$                 &  $\sim   10^{-7}$                 &  $\sim  10^{-6}$             &  $\sim  10^{-7}$  \\
  \bottomrule
 \end{tabular}
 \caption{Branching ratios for $D^0\to\pi^+\pi^- l^+ l^-$ and $D^0\to K^+K^- l^+ l^-$ from data, LHCb  \cite{Aaij:2017iyr} ($l=\mu$) and BESIII  \cite{Ablikim:2018gro} ($l=e$), our evaluation, resonant and non-resonant, and \cite{Cappiello:2012vg}. 
 Upper limits are at 90\% CL.
 $^\dagger$Statistical and systematic uncertainties are added in quadrature.}
 \label{tab:br}
\end{table}

\begin{figure}
\begin{center}
\includegraphics[width=0.45\textwidth]{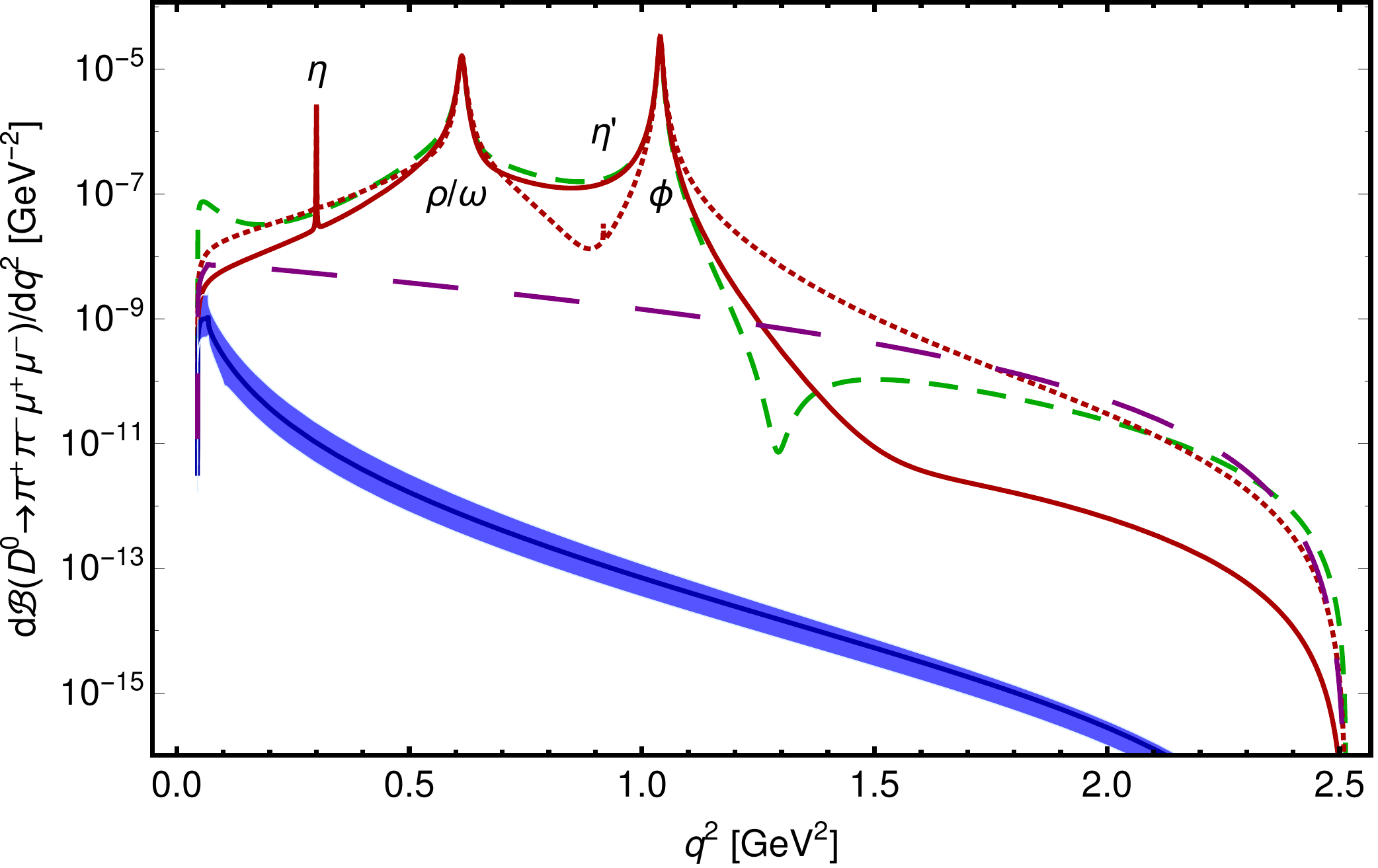}\qquad
\includegraphics[width=0.45\textwidth]{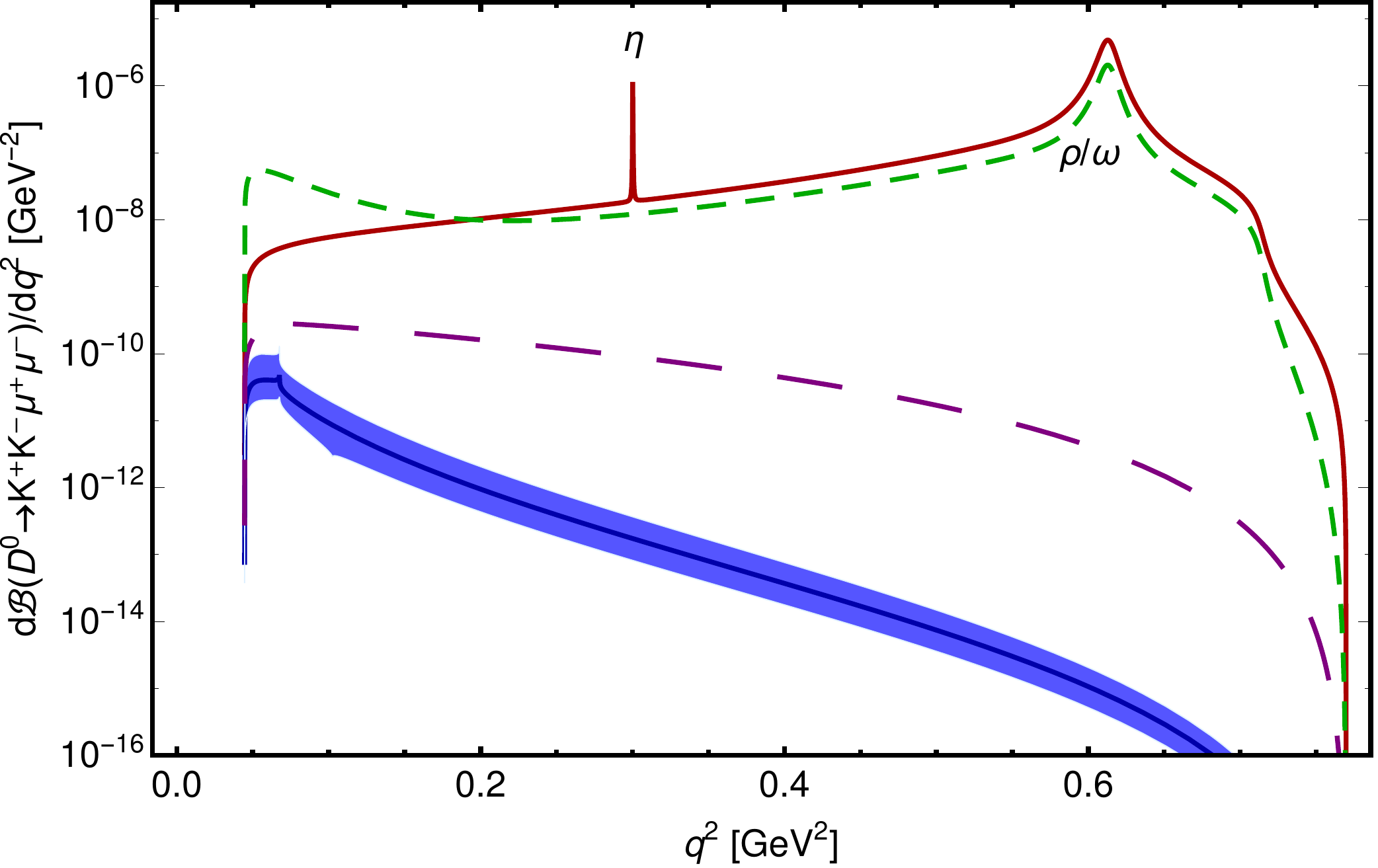}
\caption{The differential  branching ratio $d{\cal{B}}(D^0 \to \pi^+ \pi^- \mu^+ \mu^-)/dq^2$ (left) and $d{\cal{B}}(D^0 \to K^+ K^- \mu^+ \mu^-)/dq^2$ (right) in the SM for central values of input. The lowest curve (blue solid) corresponds to the non-resonant prediction including uncertainties from $m_c/\sqrt{2}\le\mu\le\sqrt{2}m_c$ represented by the band.
The long-dashed purple curve illustrates the impact of $C_9^{\rm BSM}=1$ on the non-resonant distribution.
The resonance curves are our evaluation for $\delta_\rho- \delta_\phi=\pi$ (red solid), as from $SU(3)_F$, and $\delta_\rho- \delta_\phi=0$ (red dotted) to illustrate uncertainties related to strong phases, compared to 
the model \cite{Cappiello:2012vg} (green, dashed). The latter employs fixed $\delta_\rho- \delta_\phi=\pi$ and the relative sign between the $\rho$ and the $\omega$ is as in (\ref{eq:BW}), see footnote \ref{foot:sign}. 
}
\label{fig:q2}
\end{center}
\end{figure}

In table  \ref{tab:br} branching ratio data on $D^0 \to \pi^+ \pi^- l^+ l^- $ and $  D^0 \to K^+ K^- l^+ l^-$ decays from LHCb  \cite{Aaij:2017iyr}  and BESIII  \cite{Ablikim:2018gro} are shown, together  with 
our evaluation for resonant and non-resonant branching ratios, and the  predictions from  \cite{Cappiello:2012vg}~\footnote{\label{foot:sign}There is a sign error  in eq.~(25) of   \cite{Cappiello:2012vg}: the relative sign between the $\rho$ and the $\omega$ contributions from isospin must be negative, as in our (\ref{eq:BW}). We thank
Giancarlo D'Ambrosio for confirmation.}. 
In figure  \ref{fig:q2} we show  the differential branching ratio  $d{\cal{B}}/dq^2$ for $\delta_\rho- \delta_\phi=\pi$ (red solid curve) and $\delta_\rho- \delta_\phi=0$ (red dotted curve).
The $\rho/\omega $-$\phi$ interference matters in the regions around the resonances. The $\eta^{(\prime)}$ contributions are subleading.
The purely non-resonant -- neither $q^2$ nor $p^2$ resonances are  included -- SM contribution (blue band) is much smaller than the resonance-induced distributions except for very low $q^2$.
This remains true with BSM couplings (long-dashed purple curve) as illustrated for  a maximal scenario $C_9^{\rm BSM}=1$ (\ref{eq:mubounds}).
We learn that, unlike presently in $D \to \pi \mu^+ \mu^- $ decays, in the branching ratio of $D \to \pi \pi \mu^+ \mu^- $ decays and  with  non-resonant form factors  (\ref{eq:nrFF}) there is no room left to probe BSM physics in the high $q^2$ region above the $\phi$.
In figure  \ref{fig:q2} we 
 also show the prediction by   \cite{Cappiello:2012vg} (green dashed curve).
The rise of the branching ratio at very low $q^2$ in \cite{Cappiello:2012vg} is due to the onset of  bremsstrahlung, computed using  an extrapolation of Low's theorem \cite{Low:1954kd},
an effect which will be more pronounced for electrons as lower values of $q^2$ can be accessed.
We recall that the soft photon approximation holds for photon energies up to $m^2_P/E_P$ \cite{DelDuca:1990gz}, $P=\pi, K$,  which limits its controlled use to $q^2 \lesssim 0.1 \, \mbox{GeV}^2$ 
in  $D^0 \to K^+ K^- l^+ l^- $ and to $q^2 \lesssim 0.001 \, \mbox{GeV}^2$  in $D^0 \to \pi^+ \pi^- e^+ e^- $ decays.
As it is a small effect on the $D^0 \to P_1 P_2 \mu^+ \mu^- $ branching ratios and, except for the difference in phase space a lepton universal one, we
  refrain from including this effect in  our numerics. We comment on bremsstrahlung in the discussion of LNU in section \ref{sec:LNU}.

So far we discussed $D^0 \to \pi^+ \pi^- l^+ l^- $ and $  D^0 \to K^+ K^- l^+ l^-$ decays. The former is special as it is the only one from  (\ref{eq:modes})  with a  proper distribution
at high $q^2$ above the $\phi$.  The latter decay is special as it is the only mode from  (\ref{eq:modes}) which  only proceeds through the annihilation-type topology.
On the other hand, the decays $ D^+ \to  K^+ \bar K^0   l^+ l^-$ are expected to have a more pronounced non-resonant contribution in $p^2$ as the  presumably leading resonance  in $K^+ \bar K^0 $  is $a_2(1320)$, with only a small branching ratio to $K \bar K$. The rare, semileptonic 4-body $D_s$ decays are somewhere between the two $D^0$-decays, with contributions from both topologies, however, with color-enhanced annihilation at $q^2 \simeq m^2_\phi$ and $m^2_{\eta^{(\prime)}}$. 
We stress that we employ such a phenomenological description only  to obtain BSM signatures, worked out in the next section. 
The SM predictions, that is, specific observables being null tests, are independent of the resonance model.

\section{BSM signatures \label{sec:BSM}}

In this section we work out BSM signatures of SM null tests  model-independently  and in  BSM scenarios with  leptoquarks.
For null tests related to the angular observables  $I_{5-9}$
largest effects are expected from SM-BSM interference 
near the resonances $\rho/\omega$ and $\phi$. The  dependence on the semileptonic $|\Delta c| =|\Delta u|=1$ coefficients can be taken from (\ref{eq:Iope}), (\ref{eq:BSM-dep}).

In section \ref{sec:I567} and  \ref{sec:CP}  we study the angular null tests $I_{5,6,7}$ and CP asymmetries, respectively.
In section \ref{sec:LNU} we discuss ratios of dimuon to dielectron branching ratios as a probe of LNU. LFV branching ratios are worked out in section \ref{sec:LFV}.

\subsection{Angular null tests \texorpdfstring{$I_{5,6,7}$}{I567} \label{sec:I567}}

We define integrated null test observables, normalized to the $D \to P_1 P_2  l^+ l^-$ decay rate $\Gamma$, 
\begin{align}
 \langle I_6 \rangle (q^2) & = \frac{1}{\Gamma} \int_{4 m_\pi^2}^{(m_D-\sqrt{q^2})^2} d p^2 \int_{-1}^{+1} d \cos \theta_P I_6(q^2,p^2, \cos \theta_P) \, , \label{eq:I6_ave}\\
 \langle I_{5,7} \rangle (q^2) & = \frac{1}{\Gamma} \int_{4 m_\pi^2}^{(m_D-\sqrt{q^2})^2} d p^2  \left[ \int_{0}^{+1} d \cos \theta_P  -  \int_{-1}^{0} d \cos \theta_P  \right] I_{5,7}(q^2,p^2, \cos \theta_P) \, .\label{eq:I5I7_ave}
\end{align}
We calculate  $\Gamma$ from integrating (\ref{eq:dq2p2cth}) over the full phase space.

\begin{figure}
\begin{center}
\includegraphics[width=0.45\textwidth]{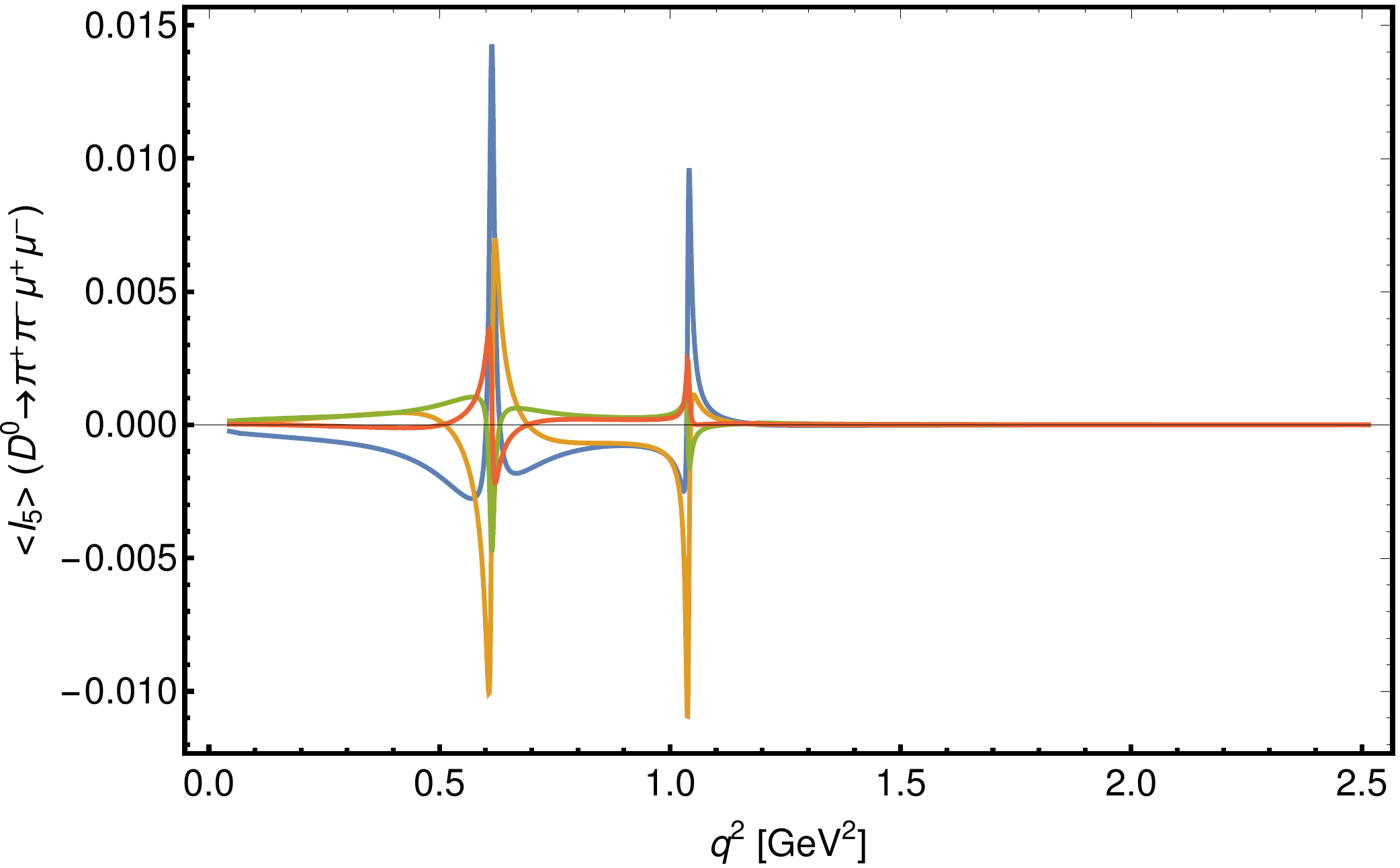}\qquad
\includegraphics[width=0.45\textwidth]{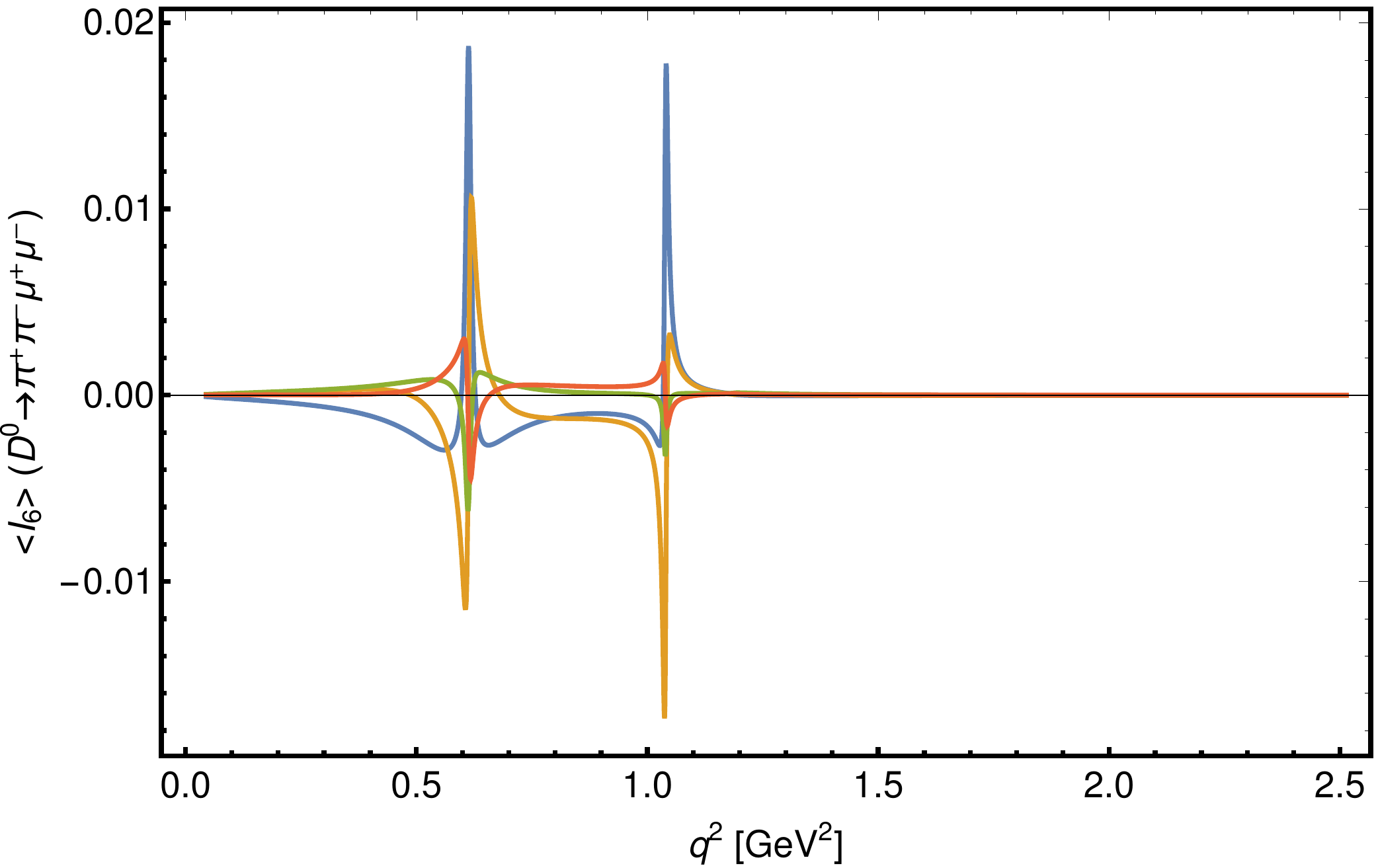}\\[1em]
\includegraphics[width=0.6\textwidth]{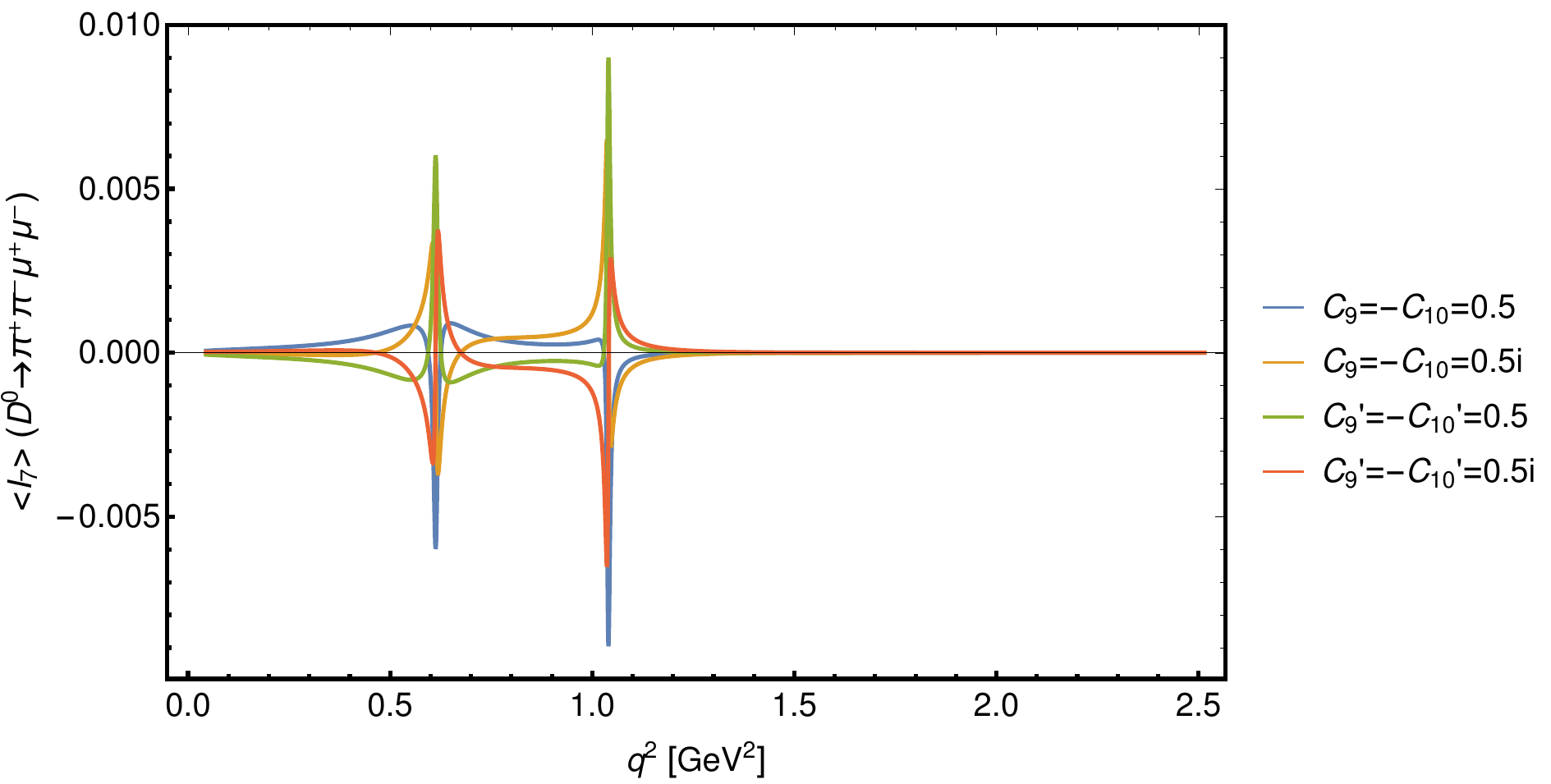}
\caption{Angular observables $\langle I_{5,6,7}\rangle$ integrated over $p^2$, see (\ref{eq:I6_ave}), (\ref{eq:I5I7_ave}), for $D^0 \to \pi^+ \pi^- \mu^+\mu^-$ normalized to $\Gamma (D^0 \to \pi^+ \pi^- \mu^+ \mu^-)$  for  $C_9^{(\prime)}=-C_{10}^{(\prime)}=0.5$, $C_9^{(\prime)}=-C_{10}^{(\prime)}=0.5i$ and relative strong phase  $\delta_\rho- \delta_\phi=\pi$.}
\label{fig:BSMq2}
\end{center}
\end{figure}

We show the integrated $I_{5,6,7}$ as a function of $q^2$ in figure \ref{fig:BSMq2} for four BSM benchmarks $C_9^{(\prime)}=-C_{10}^{(\prime)}=0.5$ and $C_9^{(\prime)}=-C_{10}^{(\prime)}=0.5i$.  
The curves for $C_9^{(\prime)}=+C_{10}^{(\prime)}=0.5$ and $C_9^{(\prime)}=+C_{10}^{(\prime)}=0.5i$ can be obtained by flipping the signs of the $\langle I_{5,6,7}\rangle$, see (\ref{eq:Iope}), (\ref{eq:BSM-dep}). The latter also explain why $I_5$ and $I_6$ have similar BSM-sensitivity and why $I_7$ is different.
As anticipated, the effects  are largest where the SM contribution peaks, around the $\rho/\omega$ and the $\phi$ resonances. The shape between the resonances depends  on their relative strong phase, shown here for   $\delta_\rho- \delta_\phi=\pi$. The effect of $\delta_\rho- \delta_\phi=0$ is a reflection of the $\phi$-peak at the $x$-axes. Our findings for the  magnitude of $\langle I_6 \rangle$ are consistent
with \cite{Cappiello:2012vg}~\footnote{Note, $\theta_l$ is defined  in \cite{Cappiello:2012vg} with respect to the positively charged lepton, whereas we use the negatively charged one. It follows that $A_{\rm FB}^{[12]}=- 2 \langle I_6 \rangle$. }.

\subsection{CP asymmetries without tagging \label{sec:CP}}

The CP asymmetries corresponding to  the CP-odd angular coefficients $I_k$, $k=5,6,8,9$  are defined as \cite{Bobeth:2008ij}
\begin{align}
A_k  = 2 \frac{I_k-\bar I_k}{\Gamma+\bar \Gamma}= \frac{ I_k-\bar I_k   }{\Gamma_{ave}} \, , 
\end{align}
where $\Gamma_{ave}$ corresponds to the CP-averaged decay rate.
The observables $I_8$ and $I_9$ can be obtained from the angular distribution (\ref{eq:full}), for instance, as follows
\begin{align}
I_8  & = \ \frac{3 \pi}{8}   \left[ \int_0^{\pi}  d  \phi  - \int_{\pi}^{2 \pi}  d \phi  \right]  \left[  \int_0^1  d\cos\theta_l  - \int_{-1}^0  d\cos\theta_l  \right] \frac{d^5\Gamma }{dq^2dp^2d\cos\theta_{P_1} d\cos\theta_l d \phi} \, , \\
I_9  & = \  \frac{3 \pi}{8} \left[ \int_{0}^{\pi/2}  d  \phi  - \int_{\pi/2}^{ \pi}  
d \phi +  \int_{\pi}^{3 \pi/2}  d  \phi  - \int_{3 \pi/2}^{ 2 \pi}  d \phi   \right]  \frac{d^4\Gamma }{dq^2dp^2d\cos\theta_{P_1}d \phi}   \, . 
\end{align}
$I_{5,6}$ are given in (\ref{eq:i5}), (\ref{eq:i6}).

We define  the integrated angular coefficients $\langle I_{8} \rangle$ analogous to $\langle I_{5,7} \rangle $, (\ref{eq:I5I7_ave}),  and  $\langle I_{9} \rangle$ 
analogous to $\langle I_{6} \rangle$, (\ref{eq:I6_ave}).
From here we obtain  the integrated CP asymmetries  $\langle A_k \rangle =(\langle I_{k} \rangle  -\langle \bar I_{k} \rangle )/\Gamma_{ave}$. Numerical values for high $q^2$, $q_\text{min}^2=(1.1\,\text{GeV})^2$ in BSM-benchmarks are  given in table \ref{tab:CPasymmetries}. To obtain the ranges given we varied strong phases and explicitly verified that the sign of $C_9$ in the first and $C_9^\prime$ in the second case does not matter, in agreement with (\ref{eq:BSM-dep}). 
In the analysis of the  CP asymmetries  in  (\ref{eq:BW}) we effectively take into account the CKM factors $V_{cd}^*V_{ud}$ and $V_{cs}^*V_{us}$ for the $\rho/\omega$ and $\phi$, respectively.
The SM predictions for $\langle A_{8,9}\rangle^\text{SM}$ at high $q^2$ are  below the permille level,
and zero for  $\langle A_{5,6}\rangle^\text{SM}$  due to the GIM-mechanism,  $C_{10}^\text{SM}=0$.
CP-asymmetries integrated over the full $q^2$ region are at most permille level in BSM models, and smaller in the SM.

\begin{table}
 \centering
 \begin{tabular}{lcc}
  \toprule
                         &  $C_9=-C_{10}=\pm0.5i$ \quad  & \quad  $C_9'=-C_{10}'=\pm0.5i$  \\
  \midrule
  $\langle A_5 \rangle$  &  $[-0.04,0.04]$  &  $[-0.03,0.03]$  \\
  $\langle A_6 \rangle$  &  $[-0.06,0.05]$  &  $[-0.06,0.06]$  \\
  $\langle A_8 \rangle$  &  $[-0.02,0.02]$  &  $[-0.02,0.02]$  \\
  $\langle A_9 \rangle$  &  $[-0.03,0.03]$  &  $[-0.03,0.03]$  \\
  \bottomrule
   \end{tabular}
 \caption{Ranges for the high $q^2$, $q_\text{min}^2=(1.1\,\text{GeV})^2$, integrated CP asymmetries $\langle A_i\rangle$ for $D^0\to\pi^+\pi^-\mu^+\mu^-$  decays for different BSM benchmarks, varying strong phases.}
 \label{tab:CPasymmetries}
\end{table}

\subsection{Testing lepton universality \label{sec:LNU}}

LNU-ratios  in semileptonic decays \cite{Hiller:2003js,Das:2014sra,Fajfer:2015mia}
\begin{align} \label{eq:LNU}
R^D_{P_1 P_2}=\frac{  \int_{q^2_{\rm min}}^{q^2_{\rm max}}d {\cal{B}}/d q^2 (D \to P_1 P_2 \mu^+ \mu^-)}{  \int_{q^2_{\rm min}}^{q^2_{\rm max}}d      {\cal{B}} /dq^2 (D \to P_1 P_2 e^+ e^-)} \, , 
\end{align}
{\it with the same cuts} in the dielectron and dimuon measurement  provide yet another null test of the SM in charm as  $R^D_{P_1 P_2}|_{\rm SM} \simeq 1$.
Phase space corrections of the order $m_\mu^2/m_c^2$ amount to percent level effects. Electromagnetic effects are another source of non-universality, and expected at order $\alpha_{em}/(4 \pi) \times logarithms$, parametrically suppressed \cite{Bobeth:2008ij,Huber:2005ig,Bordone:2016gaq}.  A detailed calculation is beyond the scope of this work.
Within  the SM we obtain for $q^2_{\rm min}=4 m_\mu^2$ and $q^2_{\rm max}=(m_D - m_{P_1} - m_{P_2})^2$ 
\begin{align} \label{eq:RD-SM}
R_{\pi \pi}^{D \, \rm SM}=1.00 \pm {\cal{O}}(\%)  \, , \quad \quad R_{KK}^{D \, \rm SM}=1.00 \pm {\cal{O}}(\%) \, .
\end{align}
Beyond the SM,  $R_{\pi \pi}^D$  can be modified significantly. Varying strong phases and Wilson coefficients $C_{9,10}^{(\prime)}$ one at a time within allowed ranges  (\ref{eq:mubounds}), 
we obtain $R_{\pi \pi}^D|_\text{BSM}\in[0.85,0.99]$  and $R_{K K}^D|_\text{BSM}\in[0.94,0.97]$. The latter is barely  distinguishable from (\ref{eq:RD-SM}), 
as well as $R_{\pi \pi}^D$ and $R_{K K }^D$  in leptoquark models, {\it e.g.}, \cite{Fajfer:2015mia,deBoer:2015boa}.
It is advantageous to consider the LNU-ratios in bins with a smaller SM contribution to increase the BSM sensitivity. For $\pi \pi$, this is, for instance, the high $q^2$ region above the $\phi$, $q^2_{\rm min}=(1.1 \, \mbox{GeV})^2$, as in \cite{Aaij:2017iyr}, and with  the SM prediction (\ref{eq:RD-SM}) intact.
Here, in  this high $q^2$ bin,  leptoquark effects are within
$R_{\pi \pi}^D|_\text{LQ}^{\text{high}\,q^2}\in[0.7,4.4]$, consistent with related sizable SM deviations in $D \to \pi l^+l^-$ decays at high $q^2$ \cite{Fajfer:2015mia}. 
Such sizable deviations from universality are possible for the 
scalar and vector $SU(2)_L$-singlet and doublet representations $S_{1,2}$, $\tilde V_{1,2}$, respectively, which  escape kaon bounds  because there is no coupling to quark doublets \cite{deBoer:2015boa}. 
The other leptoquark representations  give SM-like values for $R^D_{P_1 P_2}$.

For $D^0 \to K^+ K^- l^+ l^-$  decays we investigate possibilities to enhance the BSM sensitivity by lowering $q^2_{\rm max}$.  This increases the sensitivity to lepton mass effects such that
(\ref{eq:RD-SM}) does not hold anymore.
We find, even when simultaneously increasing $q^2_{\rm min}$, that 
leptoquark-induced LNU cannot be unambiguously distinguished from the SM in the $KK$ mode. The long-distance dominance of the branching ratio even with BSM contributions is also manifest from  figure  \ref{fig:q2}.
For instance, below the $\eta$, for $q^2_{\rm max}=(0.525 \, \mbox{GeV})^2$  \cite{Aaij:2017iyr}, we find
that $R_{K K}^D$ in leptoquark models is within  the ballpark of  the SM prediction, $R_{KK}^{D \, \rm SM}=0.83 \pm {\cal{O}}(\%)$. On the other hand, model-independently $R_{K K}^D$ can be suppressed relative to the SM,  $R_{KK}^D|_\text{BSM}^{<\eta}\in[0.60,0.87]$.

While data on muons \cite{Aaij:2017iyr}  and electrons  \cite{Ablikim:2018gro}  exist for $D^0 \to \pi^+ \pi^- l^+ l^-$ and $D^0 \to K^+ K^- l^+ l^-$ decays, see  table  \ref{tab:br}, unfortunately, this  does not permit
to compute the respective clean LNU-ratios (\ref{eq:LNU})  due to   incompatible $q^2$-cuts employed by the two experiments. In particular, BESIII  included 
$q^2$-regions not accessible with dimuons  and vetoed  the $\phi \to e^+e^-$ region. We recommend to  give dielectron results  for $q^2$ values  above the dimuon threshold to allow for 
a measurement of $R^D_{P_1 P_2}$ (\ref{eq:LNU}).
Naive ratios of the  branching ratio measurements  \cite{Aaij:2017iyr,Ablikim:2018gro} given in table  \ref{tab:br}  result in lower limits,
\begin{align}
\bar R^D_{\pi^+ \pi^-} \gtrsim  0.1  \, , \quad \quad  \bar R^D_{K^+ K^-}  \gtrsim 0.01 \, ,
\end{align}
whose respective SM predictions   are, due to the different $q^2$-cuts,  subject to sizable hadronic uncertainties. 
Using the same cuts as in the BESIII analysis  -- none on the dielectron invariant mass squared except for  excluding the region $[0.935, 1.053]$~GeV  \cite{Ablikim:2018gro} --  we find  in the model
of \cite{Cappiello:2012vg}  $\bar R^{D \, \rm SM}_{\pi^+ \pi^-} \simeq  0.9$ and $\bar R^{D \,  \rm SM}_{K^+ K^-}  \simeq0.1$, about an order of magnitude away from the data.
The smallness of the ratio $\bar R^{D \, \rm SM}_{K^+K^-}$  follows from the bremsstrahlung enhancement for electrons.
A similar effect is present in the $\pi \pi$ mode, however, here it is lifted by the contribution of the $\phi$ in the dimuon mode.
The main difference between our resonance model and \cite{Cappiello:2012vg}  is, besides the use of on-peak data  (\ref{eq:rhobin})-(\ref{eq:rhobinKK}),  the inclusion of  bremsstrahlung effects
at very low $q^2$, subject to systematic uncertainties as briefly discussed in section
\ref{sec:resonances}.
Lepton mass corrections in our model are small such that the main difference between electrons and muons is due to  the vetoed  $\phi$ in the denominator, $\bar R^{D \, \rm SM}_{\pi^+ \pi^-} \sim 2$,
and $\bar R^{D \, \rm SM}_{K^+ K^-} \sim 1$.
A measurement with identical cuts (\ref{eq:LNU})  would avoid this model-dependence.

\subsection{LFV \label{sec:LFV}}

We work out predictions for LFV branching ratios $D^0 \to  \pi^+ \pi^- e^\pm \mu^\mp$ and $D^0 \to  K^+ K^- e^\pm \mu^\mp$, which vanish in the SM.
Integrating the non-resonant distributions over the full $q^2$-range, and using the constraints discussed in section \ref{sec:weak}   we find model-independently
and in leptoquark models, following \cite{deBoer:2015boa},
\begin{align}
{\cal{B}} (D^0 \to  \pi^+ \pi^- e^\pm \mu^\mp) \lesssim 10^{-7} \, , \quad {\cal{B}} (D^0 \to  K^+ K^- e^\pm \mu^\mp) \lesssim 10^{-9}  \, . 
\end{align}

\section{Conclusions \label{sec:con}}

The SM angular distribution in semileptonic 4-body $D$-decays is considerably simpler than in  $B$-decays   because  of  long-distance dominance in charm.
The latter implies P-conservation and equal  chirality of the
lepton currents. As a result, the angular coefficients $I_{5,6,7}$ are null tests of the SM.
BSM-contributions to the axial-vector coupling, $C_{10}^{(\prime)}$, can, on the other hand, induce  rates at few percent level, see figure \ref{fig:BSMq2}.

Rare semileptonic $D^0$-decays are not self-tagging and  benefit from the CP-asymmetries related to  $I_{5,6,8,9}$, which are CP-odd and do not require $D$-tagging.
Due to the smallness of $V_{cb}^* V_{ub}/(   V_{cs}^* V_{us}   )$ corresponding CP-asymmetries $A_{5,6,8,9}$ constitute null tests of the SM.
BSM-induced integrated asymmetries can reach few percent, see table \ref{tab:CPasymmetries}.

Ratios of branching fractions into muons and electrons (\ref{eq:LNU}) probe lepton universality in the up-sector and complement studies with $B$-decays.
LNU-tests in charm are presently not very constraining  as only upper limits on branching ratios of $D \to P_1 P_2 e^+ e^-$ decays  exist.  We strongly encourage
experimenters to  provide in the future data based on the same kinematic cuts for muons and electrons, enabling more powerful SM tests.

Leptonic P-invariance  and suppression of SM CP violation  holds  in the whole $(p^2,q^2)$-phase space on and off resonance peaks. Therefore,
there is no particular need for cutting on $\pi \pi$ around or outside the $\rho$, or $ll$ around $\phi$ or $\rho/\omega$ and  one can collect events from the whole phase space. Yet, 
experimental information on the otherwise  SM-dominated branching ratios with  on-resonance cuts assists tuning the hadronic model parameters.
Note, near-resonance  BSM signals in the angular observables $I_{5 -9}$ are larger due to enhanced interference with the SM, as exploited in  \cite{Fajfer:2012nr,Cappiello:2012vg} and evident in figure \ref{fig:BSMq2}.
On the other hand, deviations from lepton universality in the ratios  (\ref{eq:LNU})  are enhanced in regions where the SM-contribution is smaller, such as in the high $q^2$ region above the $\phi$ in
$D^0 \to \pi^+ \pi^- l^+ l^-$ decays, where order one BSM effects are possible.
LFV branching ratios ${\cal{B}} (D^0 \to  \pi^+ \pi^- e^\pm \mu^\mp)$  and  $ {\cal{B}} (D^0 \to  K^+ K^- e^\pm \mu^\mp)$ can reach  $10^{-7}$ and  $10^{-9}$, respectively.

\bigskip

{\bf Acknowledgements}\\
GH would like to thank the participants of the  "Towards the Ultimate Precision in Flavour Physics" (TUPIFP) workshop, held April 16-18, 2018   at Warwick U for stimulating discussions.
This work has been supported  by the
DFG Research Unit FOR 1873 ``Quark Flavour Physics and Effective Field Theories''
and by the BMBF under contract no. 05H15VKKB1.

\section{Appendix: \texorpdfstring{$D \to P_1 P_2 l^+ l^-$}{DtoP1P2ll}  matrix elements}

\subsection{\texorpdfstring{$D \to P_1 P_2 $}{DtoP1P2} form factors}

We employ the form factors from HH$\chi$PT \cite{Lee:1992ih}
\begin{align} \label{eq:nrFF}
 &w_\pm=\pm\frac{\hat gf_D}{2f_{P_1}^2}\frac{m_D}{v\cdot p_{P_1}+\Delta}\,,\quad h=\frac{\hat g^2f_D}{2f_{P_1}^2}\frac1{(v\cdot p_{P_1}+\Delta)(v\cdot p+\Delta)}\,,
\end{align}
with input $\Delta=(m_{{D^*}^0}-m_{D^0})=0.1421\,\text{GeV}$,   $f_D=0.21\,\text{GeV}$, $f_\pi=0.13\,\text{GeV}$, $f_K=0.156\,\text{GeV}$, $ \hat g=0.570\pm0.006$ \cite{Lees:2013zna}, $v\cdot p_{P_1}=((m_D^2-q^2+p^2)-\sqrt{\lambda(m_D^2,q^2,p^2)(1-4m_{P_1}^2/p^2)}\cos\theta_{P_1})/(4m_D)$ and $v\cdot p=(m_D^2-q^2+p^2)/(2m_D)$.

\subsection{Resonance amplitudes}

The transversity form factors for the  contributions from resonances $R$ with spin $J_R$ read \cite{Das:2014sra}
\begin{align} \label{eq:fullformfactor}
{\cal F}_0 & \equiv {\cal F}_0(q^2, p^2, \cos \theta_{P_1}) \simeq \sum_R P^{0}_{J_R} (\cos\theta_{P_1}) \cdot F_{0 J_R}(q^2, p^2)  \, ,   \\
{\cal F}_i & \equiv  {\cal F}_i(q^2, p^2, \cos \theta_{P_1}) \simeq \sum_R \frac{P^{1}_{J_R} (\cos\theta_{P_1}) }{\sin \theta_{P_1}}  \cdot F_{iJ_R}(q^2, p^2)  \, , \quad i= \parallel, \perp \, , \nonumber
\end{align}
where $P^m_{\ell}$ denote the associated Legendre polynomials, {\it e.g.}, $P_1^0(\cos\theta_P)=\cos\theta_P$ and $P_1^1(\cos\theta_P)=-\sin\theta_P$.
For vector $V$ resonances with mass $m_V$ and width $\Gamma_V$ \cite{Das:2014sra,Das:2015pna}
\begin{align}
 &F_{0V}=-3N_V\frac{(m_D^2-m_V^2-q^2)(m_D+m_V)^2A_1(q^2)-\lambda(m_D^2,m_V^2,q^2)A_2(q^2)}{2m_V(m_D+m_V)\sqrt{q^2}}P^V\,,\\
 &F_{\parallel V}=-\frac3{\sqrt2}N_V\sqrt2(m_D+m_V)A_1(q^2)P^V\,,\\
 &F_{\perp V}=\frac3{\sqrt2}N_V\frac{\sqrt{2\lambda(m_D^2,m_V^2,q^2)}}{m_D+m_V}V(q^2) P^V\,, \label{eq:Fperp}
\end{align}
with the resonance shape $P^V$.
For the latter we employ  a Breit-Wigner parametrization \cite{delAmoSanchez:2010fd}, 
\begin{align}
 &P^V(p^2)=\sqrt{\frac{m_V\Gamma_V}\pi}\frac{p^*}{p_0^*}\frac1{p^2-m_V^2+im_V\Gamma_V(p^2)}\,,\\
 &\Gamma_V(p^2)=\Gamma_V\left(\frac{p^*}{p_0^*}\right)^3\frac{m_V}{\sqrt{p^2}}\frac{1+(r_{BW}\,p_0^*)^2}{1+(r_{BW}\,p^*)^2}\,,\\
 &p^*=\frac{\sqrt{\lambda(p^2,m_{P_1}^2,m_{P_2}^2)}}{2\sqrt{p^2}}\,,\quad p_0^*=p^*|_{p^2=m_V^2}\,,
\end{align}
which is normalized $\int d p^2 |P^V(p^2)|^2 =1$. For the  $\rho$ we use
the Blatt-Weisskopf parameter $r_{BW}=3\,\text{GeV}^{-1}$   \cite{CLEO:2011ab}.
In the normalization factor
\begin{align}
 &N_V=G_F\alpha_e\sqrt{\frac{ \beta_l q^2\sqrt{\lambda(m_D^2,p^2,q^2)}}{3(4\pi)^5m_D^3}}\,, \quad \beta_l=\sqrt{1- \frac{4 m_l^2}{q^2}} \, , 
\end{align}
we use the K\"allen-function suitable for off-resonance effects,  instead of $\lambda(m_D^2,m_V^2,q^2)$, and include an overall finite $m_l$ phase space suppression.
Form factors $A_{1,2}, V$ are provided  in \cite{Melikhov:2000yu,Wu:2006rd,Verma:2011yw}.
Following \cite{deBoer:2017que} we  employ the form factors by  \cite{Melikhov:2000yu}, parameterized as 
\begin{align}
 F(q^2)=\frac{\tilde F(0)}{1-\sigma_1\,q^2/m_{D^*}^2}\,,
\end{align}
where $\tilde F(0)=F(0)/(1-q^2/m_{D^*}^2)$ for $F=V$ and $\tilde F(0)=F(0)$ for $F=A_{1,2}$.
For $D\to\rho$ the parameters are given as
\begin{align}
 &V(0)=0.90\,,&&\sigma_1=0.46\,,\,\nonumber\\
 &A_1(0)=0.59\,,&&\sigma_1=0.50\,,\, \\
 &A_2(0)=0.49\,,&&\sigma_1=0.89\,\,.\nonumber
\end{align}
Since the modelling of the resonances itself is accompanied by large uncertainties, we neglect the form factor uncertainties in the numerical evaluations as well as differences between
$D \to \rho$ and $D \to \omega$ form factors.

For the resonance-induced $D^0 \to K^+ K^- l^+ l^-$ contribution we use  the form factors 
\begin{align}
 &F_{0\phi}=-3 N_V  \frac{(m_D^2-m_\rho^2-p^2)(m_D+m_\rho)^2A_1(p^2)-\lambda(m_D^2,m_\rho^2,p^2)A_2(p^2)}{2m_\rho(m_D+m_\rho)\sqrt{q^2}}P^\phi\,,\\
 &F_{\parallel \phi}=-\frac3{\sqrt2}N_V\sqrt2(m_D+m_\rho)A_1(p^2)P^\phi\,,\\
 &F_{\perp \phi}=\frac3{\sqrt2}N_V\frac{\sqrt{2\lambda(m_D^2,m_\rho^2,p^2)}}{m_D+m_\rho}V(p^2) P^\phi\,,
\end{align}
where $V,A_{1,2}$ are $D \to \rho$ form factors  given above. We employ a constant width (normalized)  Breit-Wigner distribution for the $\phi$-lineshape
\begin{align}
 &P^\phi(p^2)=\sqrt{\frac{m_\phi\Gamma_\phi}\pi} \frac1{p^2-m_\phi^2+im_\phi\Gamma_\phi}\,.
\end{align}

\end{document}